\def\msun{\hbox{M$_{\odot}$}}
\def\lsun{\hbox{L$_{\odot}$}}
\def\kms{\hbox{km s$^{-1}$}}

\def\lesssim{\mathrel{\hbox{\rlap{\hbox{\lower2pt\hbox{$\sim$}}}\raise2pt\hbox{$<$}}}} 
\def\hb{\hbox{H$\beta$}}
\def\ha{\hbox{H$\alpha$}}

\def\oiii{\hbox{[OIII]}}
\def\nii{\hbox{[NII]}}
\def\sii{\hbox{[SII]}}

\def\oiiihb{\hbox{[OIII]/H$\beta$}}
\def\hahb{\hbox{H$\alpha$/H$\beta$}}
\def\niiha{\hbox{[NII]/H$\alpha$}}
\def\siiha{\hbox{[SII]/H$\alpha$}}
\documentclass[useAMS,usenatbib]{mn2e}

\title[Ionized outflows in the nearest obscured quasars]{The triggering mechanism and properties of  ionized outflows in the nearest obscured quasars}
\author[Villar Mart\'\i n et al.]{M. Villar Mart\'\i n$^{1}$, B. Emonts$^{1}$,   A. Humphrey$^2$, A. Cabrera Lavers$^3$,  L. Binette$^4$ \\
$^1$Centro de Astrobiolog\'\i a (INTA-CSIC), Carretera de Ajalvir, km 4, 28850 Torrej\'on de Ardoz, Madrid, Spain.  villarmm@cab.inta-csic.es \\
$^2$Centro de Astrofisica, Universidade do Porto, Rua das Estrelas, 4150-762 Porto, Portugal \\
$^3$Instituto de Astrof'sica de Canarias, (IAC) V'a L‡ctea s/n, La Laguna, Tenerife, Spain\\
$^4$Instituto de Astronom\'\i a, Universidad Nacional Aut\'onomo de M\'exico, Ap. 70-264, 04510 M\'exico D.F., M\'exico}

\begin{document}

\date{Accepted 2014 March 05. Received 2014 March 05; in original form 2013 December 17.}

%\pagerange{\pageref{firstpage}--\pageref{lastpage}} \pubyear{2002}

\maketitle

\label{firstpage}

\begin{abstract}

We have identified ionized outflows in the narrow line region of all  but one SDSS type 2 quasars (QSO2) at $z\la$0.1 (20/21, detection rate 95\%), implying that this is a ubiquitous phenomenon in this object class also at the lowest $z$.  The outflowing gas has high densities ($n_e\ga$1000 cm$^{-3}$)  
and covers a region the size of a few kpc. This implies ionized outflow masses  $M_{outf}\sim$(0.3-2.4)$\times$10$^6$ \msun~ and mass outflow rates $\dot M<$few \msun~ yr$^{-1}$. 

The triggering mechanism of the outflows is related to the nuclear activity.
The QSO2 can be classified in two  groups according to the behavior and properties of the outflowing gas.
QSO2 in Group 1 (5/20 objects) show the most extreme turbulence, they have on average higher radio luminosities and higher  excess of radio emission.   QSO2 in Group 2 (15/20 objects) show less extreme turbulence, they have lower radio luminosities and, on average, lower or no radio excess. 

We propose that two competing 
outflow mechanisms are at work:  radio jets and accretion disk winds.  
Radio jet induced outflows are dominant in Group 1, while  disk winds dominate in Group 2.  We find that the radio jet
mode is  capable of producing  more extreme outflows. 

To test this interpretation we predict that: 1)  high resolution VLBA imaging will reveal the presence of jets in  Group 1 QSO2; 2) 
the morphology of their extended ionized nebulae must be more highly collimated and kinematically perturbed.

\end{abstract}

\section{Introduction}

Evidence for an intimate connection between supermassive black hole (SMBH) growth and the evolution of galaxies is nowadays compelling. Not only have SMBHs  been found in many galaxies with a bulge component, but correlations exist between the black hole mass and some bulge properties, such as the stellar mass and velocity dispersion (e.g. Ferrarese \& Merritt \citeyear{fer00}). The origin of this relation is still an open question, but quasar induced outflows might play a critical role. Hydrodynamical simulations show that the energy output from quasars can regulate the growth and activity of black holes and their host galaxies (e.g. di Matteo, Springel, Hernquist \citeyear{dim05}). Such models also show that  the energy released by the strong  outflows associated with major phases of accretion  expels enough gas to quench both star formation and further black hole growth. Growing observational evidence for the dramatic impact that quasar induced outflows may have on their host galaxies
 is accumulating. Such impact is likely to depend on the  luminosity of the active galactic nucleus (AGN) and to be more efficient at the highest luminosities 
(Page et al.  \citeyear{pag12}).

Studies of powerful radio galaxies show that the radio structures can induce powerful outflows (Humphrey et al. \citeyear{hum06})   
which  may have enormous energies sufficient to eject a large fraction of the gaseous content of the galaxy (Nesvadba et al.  \citeyear{nes06}, Morganti, Tadhunter \& Oosterloo \citeyear{mor05}).
However, only $\sim$10\%  of active galaxies are radio loud. How  AGN  feedback
works in radio quiet objects is still an open question.

Several mechanisms could be at work, including stellar winds, radio jet induced outflows and accretion disc winds.
Even in radio quiet quasars the presence of radio jet induced winds  cannot be discarded.  Indeed, the existence of such jets
 in many (all?) radio quiet  AGN has been proposed by different authors (e.g.  Ghisellini, Haardt \& Matt  \citeyear{ghi04},   Ho \& Peng \citeyear{ho01}, Falcke \& Biermann  \citeyear{fal95}, Malzac et al.  \citeyear{mal98}) even if in many of these systems the jet is likely to be aborted. This is supported by the VLBI
imaging of radio quiet Seyfert galaxies (Ulvestad  \citeyear{ulv03} and references therein) which revealed the presence of  mini-jets (sub parsec scale) in many of
them. Mass loss via an accretion disc wind (driven out by radiation pressure, magneto-hydrodynamic or magneto-centrifugal forces) has also been proposed (see Hamman et al. \citeyear{ham13} and references therein). 

Type 2 quasars (QSO2)
are unique objects for investigating  the way  feedback works in the most powerful radio quiet AGN.  
The active  nucleus in QSO2 is occulted by  obscuring material, which  acts like a convenient 
``natural coronograph'',
allowing a detailed study of many properties  of the surrounding medium. 
 This 
is very complex in type 1 quasars (QSO1) due to the dominant contribution of the quasar
point spread function.

QSO2 have been discovered in large quantities only in recent years. 
 In particular,  \cite{zak03} and  \cite{rey08}   have  identified $\sim$900  objects at redshift  
$z \lesssim$0.8 in the Sloan Digital Sky Survey (SDSS, York et al.  \citeyear{york00})  with the high ionization
narrow emission line spectra characteristic of type 2 AGN and  narrow
line luminosities typical of QSO1 (log(L\oiii/L$_{\odot}$)$>$8.3). Based on a  spectroscopic study  of 13 SDSS QSO2 at 0.3$\lesssim z \lesssim$0.6, Villar-Mart\'\i n et al. (2001b, hereafter \cite{vm11b}) 
found clear evidence for  ionized outflows in the majority of objects  and argue that this is  a ubiquitous phenomenon in  the nuclear region ($r<$several kpc) of this object class.    \cite{liu13b} propose that the ionized outflows extend much farther, and can reach distances
$\ga$15 kpc (see also Humphrey et al. 2010).  The ionization, kinematic and morphological properties of the outflowing
gas  (\cite{vm11b}, Liu et al.  \citeyear{liu13b})  suggest that the {\it ionized} outflows are preferentially induced by AGN related 
processes. 

\cite{lal10} studied the radio properties of 59 SDSS QSO2 at $z\ga$0.3. The detection rate of their survey is 59\% (35/59).
They find that  15\%$\pm$5\%  can be considered radio loud, according to the  radio-to-[OIII]$\lambda$5007 luminosity ratios, while     the vast majority
of their detected sources   fall in a region intermediate between those traditionally occupied by
radio loud and radio quiet quasars. They detect a high fraction (75\%) of compact cores, which
confine the radio emission to typical physical diameters of 5 kpc or less. Thus, both the radio jet and disc wind modes are possibly present.

We present in this paper the results of our spectral analysis of the nearest SDSS QSO2 (21 objects at $z\la$0.1 from Reyes et al. \citeyear{rey08}). Based on the spectral decomposition of the
most important optical emission lines using the SDSS spectra, we search for signatures of ionized outflows. By isolating the emission from the quiescent and turbulent (outflow) components,  we compare their kinematic properties
and line ratios and characterize how the  outflows alter the properties of the ambient gas. We also investigate whether the properties of  the turbulent gas 
depend on radio loudness. Our goal is to understand whether the radio structures play a role in triggering the outflows.
By studying the outflows in the most nearby, luminous radio quiet quasars,
 we hope to shed light on feedback in the most distant quasars.

We assume
$\Omega_{\Lambda}$=0.7, $\Omega_{\rm M}$=0.3, H$_0$=71 km s$^{-1}$ Mpc$^{-1}$.  At $z=$0.1, 1$\arcsec$ corresponds  to 1.8 kpc.

\section{The sample}
 
We concentrate our analysis on the most nearby QSO2 ($z\la$0.1) from the SDSS database (Reyes et al. \citeyear{rey08}).
We optimize in this way the chances of detecting and isolating the often very faint signatures of the ionized outflows in different emission lines  (e.g. faint broad wings, which are often detected
at $\la$20\% level of the main emission line peak). We moreover ensure that the main emission lines of interest to us  (from \hb~to [SII]$\lambda\lambda$6716,6731) are within the observed spectral range for all objects.

There are 25  QSO2 at $z\le$0.1 and with log(L\oiii/L$_{\odot}$)$>$8.3
in the catalogue of \cite{rey08}.  To facilitate the analysis
and interpretation, we have excluded  4 objects  which show evidence of 
an underlying broad component in the Balmer lines indicative of contaminating line emission from a broad line region. 
Our final sample consists of 21 objects (Table \,\ref{tab:objects}). The $z$ distribution has median, average and standard deviation values  0.087, 0.083 and 0.017 respectively.

The optical morphologies of the systems as revealed by the SDSS images are very diverse: complex  mergers, disk/spiral galaxies and ellipticals (Fig. \ref{fig:gallery}). 

\begin{figure*}
\includegraphics{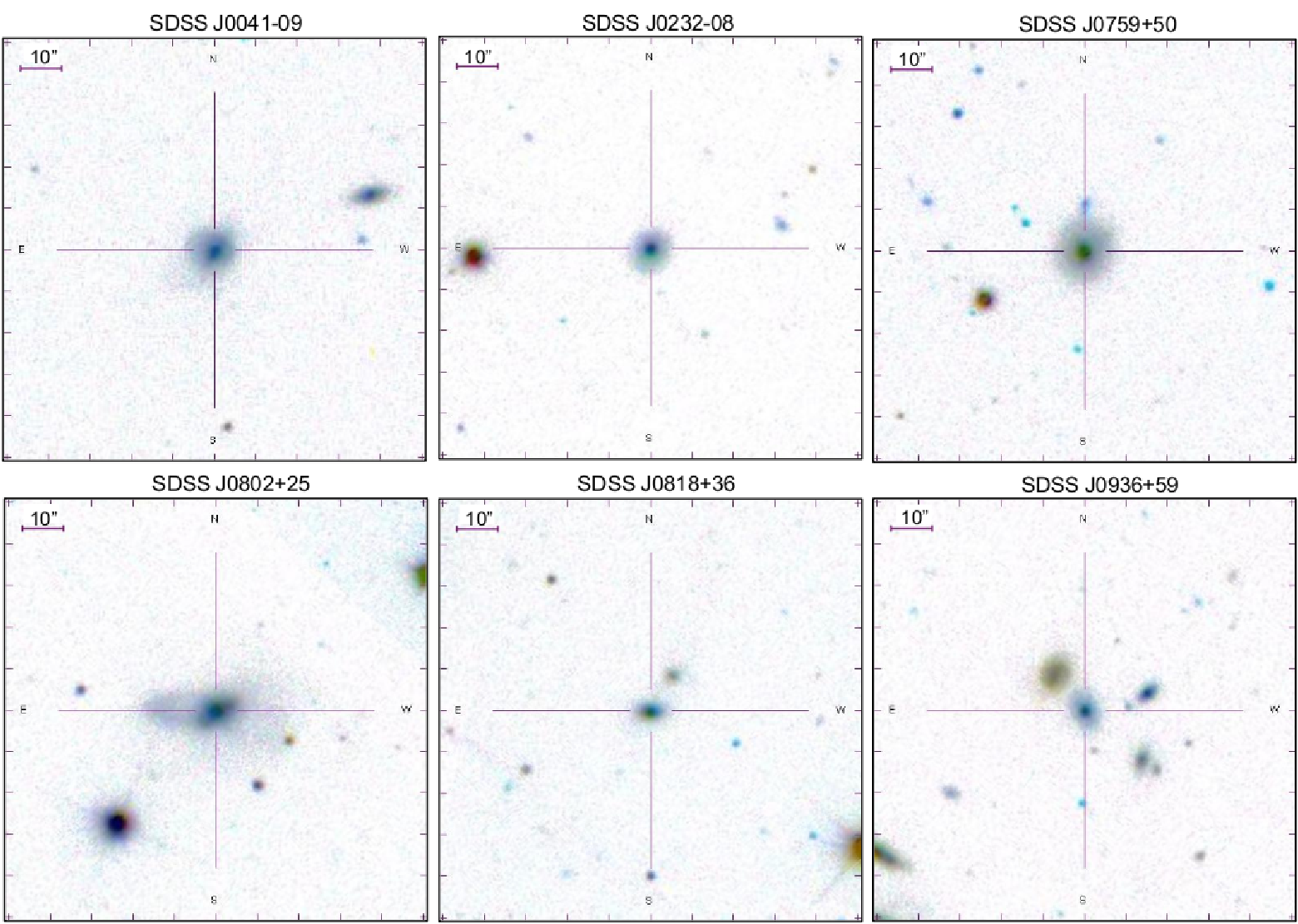}
\vspace{4.5in}
\includegraphics{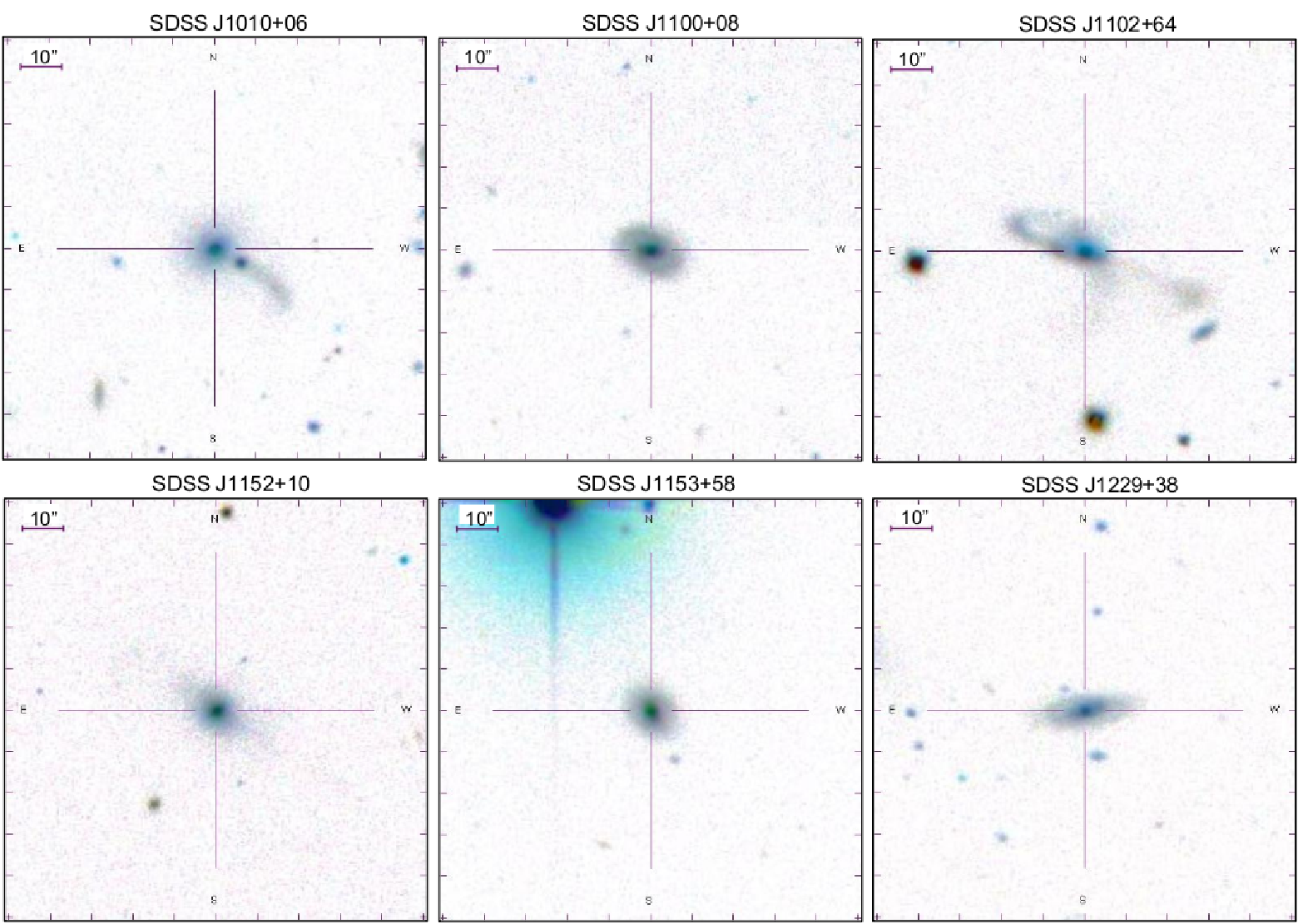}
\vspace{4.5in}
\caption{SDSS images of the QSO2 sample.}
\label{fig:gallery}
\end{figure*}

\begin{figure*}
\includegraphics{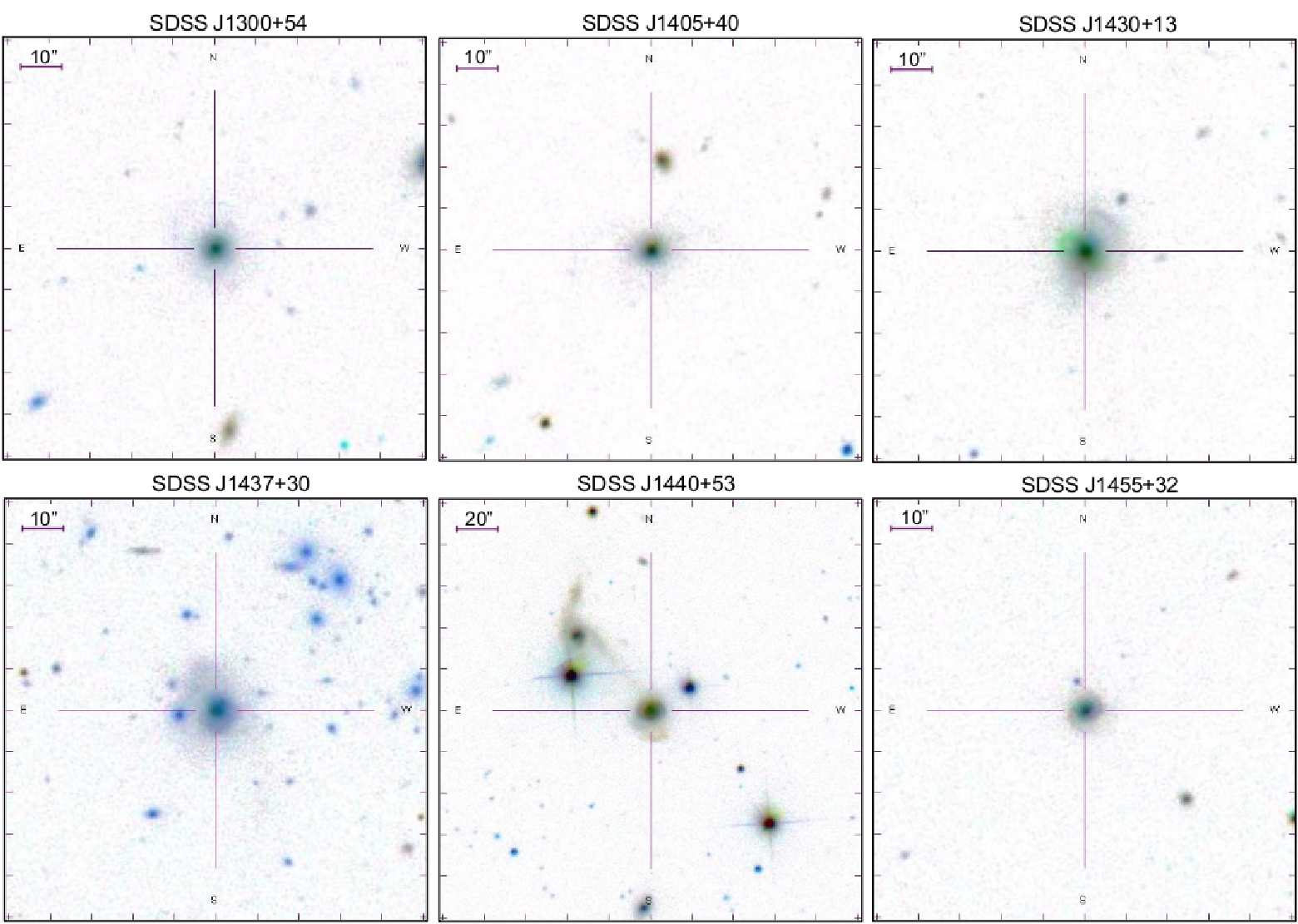}
\vspace{4.5in}
\includegraphics{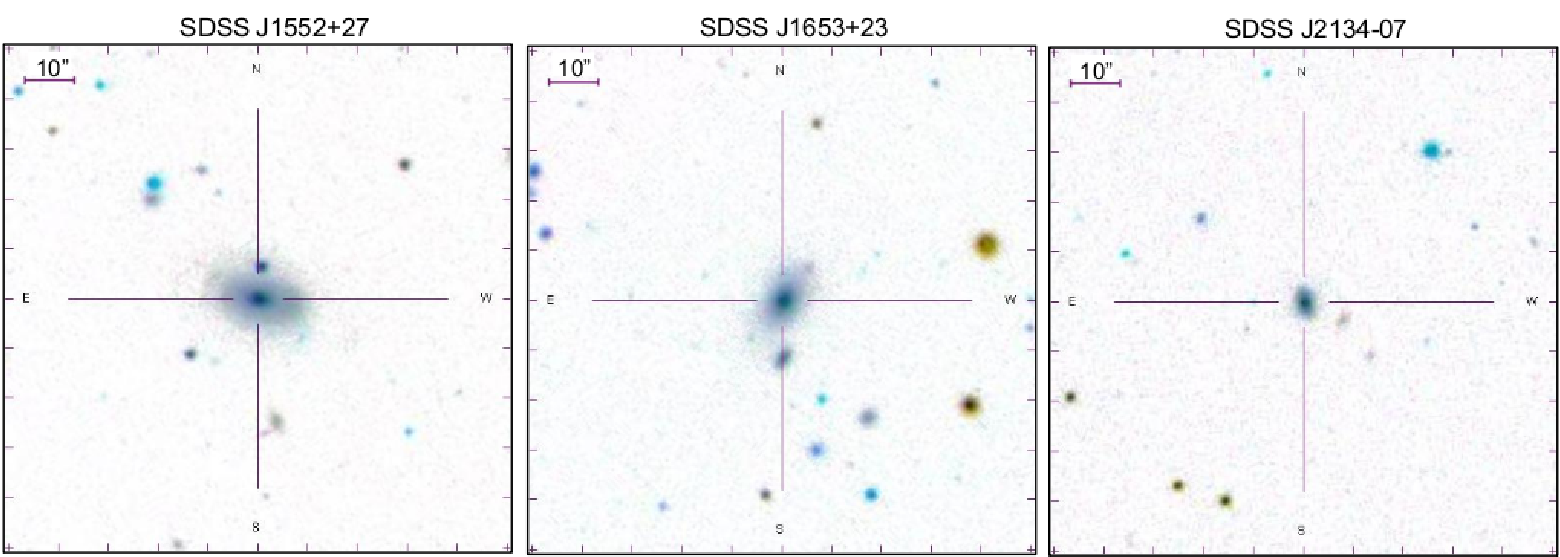}
\vspace{2.3in}
\contcaption{}
\label{fig:gallery2}
\end{figure*}

\begin{figure*}
\includegraphics{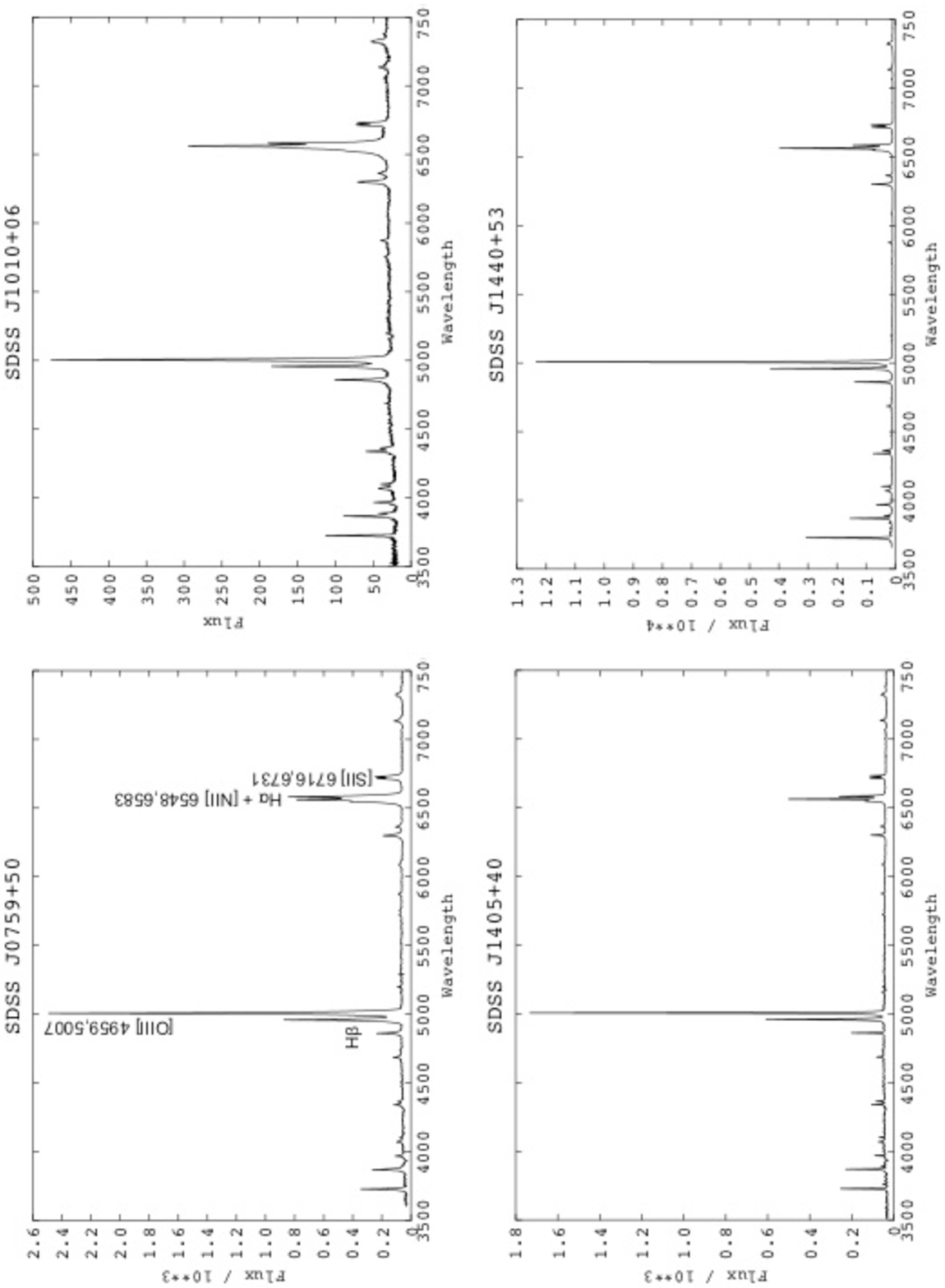}
\vspace{4.0in}
\caption{SDSS spectra of several QSO2 in our sample. Flux is given in units of $\times$10$^{-17}$ (SDSS J1010+06), $\times$10$^{-14}$ (SDSS J0759+50, SDSS J1405+40) or $\times$10$^{-13}$ (SDSS J1440+43) erg s$^{-1}$ cm$^{-2}$ \AA$^{-1}$. The lines of interest for our analysis are indicated in the first panel.}
\label{fig:spectra}
\end{figure*}

All spectra present very  strong emission lines. Several examples are shown in Fig. \ref{fig:spectra}.

\begin{table*}
\centering
\begin{tabular}{lllllll} 
\hline
Object & RA & Dec &  $z$ &  L[OIII] & $S_{20 \rm cm}$ & $S_{\rm 60\,\mu m}$  \\
&	& & &   &   mJy & Jy  \\ \hline
SDSS J0041-09 & 00 41 13.754 &	-09 52 31.66	 & 0.095 &  8.33 & $\le$0.5  & $\le$0.20    \\
SDSS J0232-08  & 02 32 24.245	& -08 11 40.23 & 0.100 &  8.60 & 4.1 & $\le$0.20  \\
SDSS J0759+50 &  07 59 40.961 &	+50 50 24.05 & 0.054 &  8.83 & 45.6 & 0.71 \\ 
SDSS J0802+25  &  08 02 52.927	& +25 52 55.59 & 0.081 & 8.86 & 29.4 & 0.51 \\
SDSS J0818+36 & 08 18 42.356 &	+36 04 09.69	 & 0.076 & 8.53 & 2.2 & $\le$0.20 \\
SDSS J0936+59  & 	09 36 25.371 &	+59 24 52.70	 & 0.096 & 8.35 & 3.9 & $\le$0.20 \\
SDSS J1010+06 &  10 10 43.367 &	+06 12 01.44 & 0.098 &  8.68 & 92.1 & 0.59 \\
SDSS J1100+08 & 11 00 12.393 &	+08 46 16.37 & 0.100 & 9.20 & 58.5 & $\le$0.20 \\
SDSS J1102+64  & 11 02 13.015	& +64 59 24.84 & 0.078 & 8.45 &  45.2 & 0.95  \\
SDSS J1152+10 & 11 52 45.659	& +10 16 23.84 & 0.070 & 8.72 & 3.6 & $\le$0.20  \\
SDSS J1153+58 &  11 53 26.430	& +58 06 44.61 & 0.065  &  8.48 & 8.1 & $\le$0.20 \\
SDSS J1229+38 & 12 29 30.407	& +38 46 20.67 & 0.102 & 8.41 & 2.9  & $\le$0.20 \\
SDSS J1300+54 & 13 00 38.097 &	+54 54 36.88 & 0.088 & 8.94 & 2.2  & $\le$0.20  \\
SDSS J1405+40  & 14 05 41.210 &	+40 26 32.55	 & 0.081 &  8.78 & 16.8 & 0.26  \\
SDSS J1430+13 & 14 30 29.886 &	+13 39 12.05 & 0.085 &  9.08 &  26.4 & 0.26 \\
SDSS J1437+30 & 14 37 37.852 & 	+30 11 01.12 & 0.092 & 8.82 & 63.9 & 0.24 \\
SDSS J1440+53 & 14 40 38.097 &	+53 30 15.88 & 0.037 &  8.94 &  57.6  & $\le$0.20   \\
SDSS J1455+32 & 14 55 19.409  &	+32 26 01.82 & 0.087 & 8.64 &  2.8  & $\le$0.20   \\
SDSS J1552+27 & 15 52 25.671 &	+27 53 43.48	 & 0.074 & 8.40 & $\le$0.5  & $\le$0.20   \\
SDSS J1653+23  & 16 53 15.051 &	+23 49 42.95 &  0.103 & 9.00 &  6.9 & 0.49  \\
SDSS J2134-07 & 21 34 00.608 &	-07 49 42.69 &  0.089 & 8.34 & 4.3  & $\le$0.20  \\ 
\hline
\end{tabular}
\caption{SDSS QSO2 at $z\la$0.1 in the sample analyzed in this work. The [OIII]$\lambda$5007 luminosities are given
in log and relative to \lsun.}
\label{tab:objects}
\end{table*}

\section{Data set and analysis}

\subsection{Optical}
\label{sec:data_opt}

The data  set consists of  optical 1-dim spectra of the QSO2 sample from the SDSS. Each spectrum corresponds to an aperture defined by 
the 3" diameter fiber, or $\sim$5.5 kpc at $z=$0.1. The rest frame spectra cover in all cases  all the main emission lines from [OII]$\lambda$3727 to [SII]$\lambda$6716,6731 (Fig. \ref{fig:spectra}).  The lines of interest to us are H$\beta$, [OIII]$\lambda\lambda$4959,5007, H$\alpha$, [NII]$\lambda$6548,6583 and [SII]$\lambda\lambda$6716,6731.
They have large equivalent widths  in all quasars. Moreover, the stellar features (if detected) are comparatively weak. 
Underlying stellar absorption of the Balmer lines is expected to be negligible and therefore subtracting  the stellar continuum is not necessary. 

The spectral line profiles   were fitted  with the {\small STARLINK}  package {\small DIPSO}  following the same procedure 
described in \cite{vm11b}.  DIPSO is based on the optimization of fit coefficients, in the sense of minimizing the sum of the squares of the deviations of the fit from the spectrum data.
 The output from a completed fit consists of the optimized parameters and their errors (calculated in the linear approximation, from the error matrix).
 
\begin{figure*}
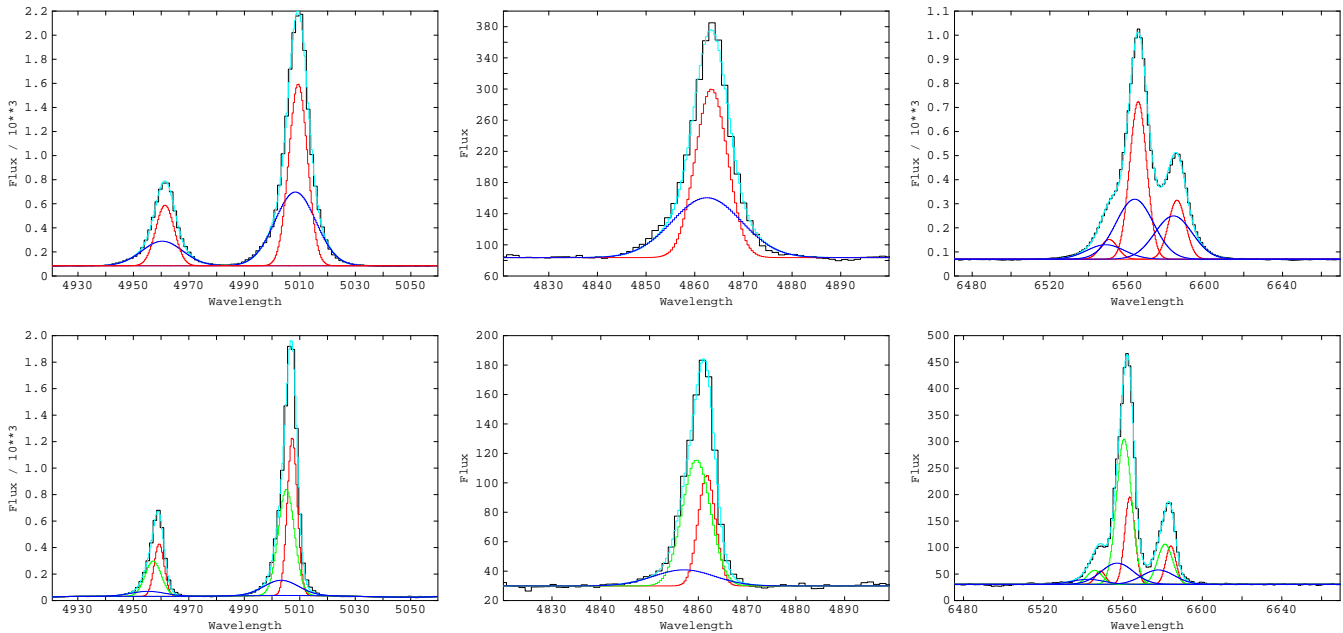

\includegraphics{fitoiii1430+13.ps}
\includegraphics{fithb1430+13.ps}
\includegraphics{fitniiha1430+13.ps}
\vspace{1.7in}
\includegraphics{fitoiii1653+23new.ps}
\includegraphics{fithb1653+23new.ps}
\includegraphics{fitniiha1653+23new.ps}
\vspace{1.7in}
\caption{Examples of fits of the main emission lines for two QSO2 in the sample:  SDSS J1653+23 (top) and SDSS J1450+53 (bottom). \oiii~ (left), \hb~(middle) and \nii+\ha ~(right). The original spectra and the fits are shown in black and cyan respectively. The same color code (blue, green or red) is used for a given kinematic component in all emission lines for a given object. Blue and red are used for the most blueshifted
and redshifted components respectively.}
\label{fig:fits}
\end{figure*}

 The [OIII] lines are in all cases the strongest and are moreover isolated (unlike H$\alpha$+[NII], which are in general severely blended). For these reasons, the \oiii~lines   were used to constrain the  kinematic substructure. The same kinematic components were forced to be present in all other emission lines with the same FWHM and separation in  \kms. For the  [OIII], [NII] and [SII] doublets the FWHM of the two doublet components were forced to be identical and the  separation in $\lambda$ was fixed with the value predicted by theory. The flux ratios [OIII]$\frac{\lambda 5007}{\lambda 4959}$ and  [NII]$\frac{\lambda 6583}{\lambda 6548}$  were  fixed to the theoretical values  2.99 and 2.88 respectively (Osterbrock \citeyear{ost89}).  
Fits  with unphysical results (e.g. kinematic components narrower
than the instrumental resolution) were rejected. Several fit examples  are shown in Fig. \ref{fig:fits}.

\begin{figure}
\includegraphics{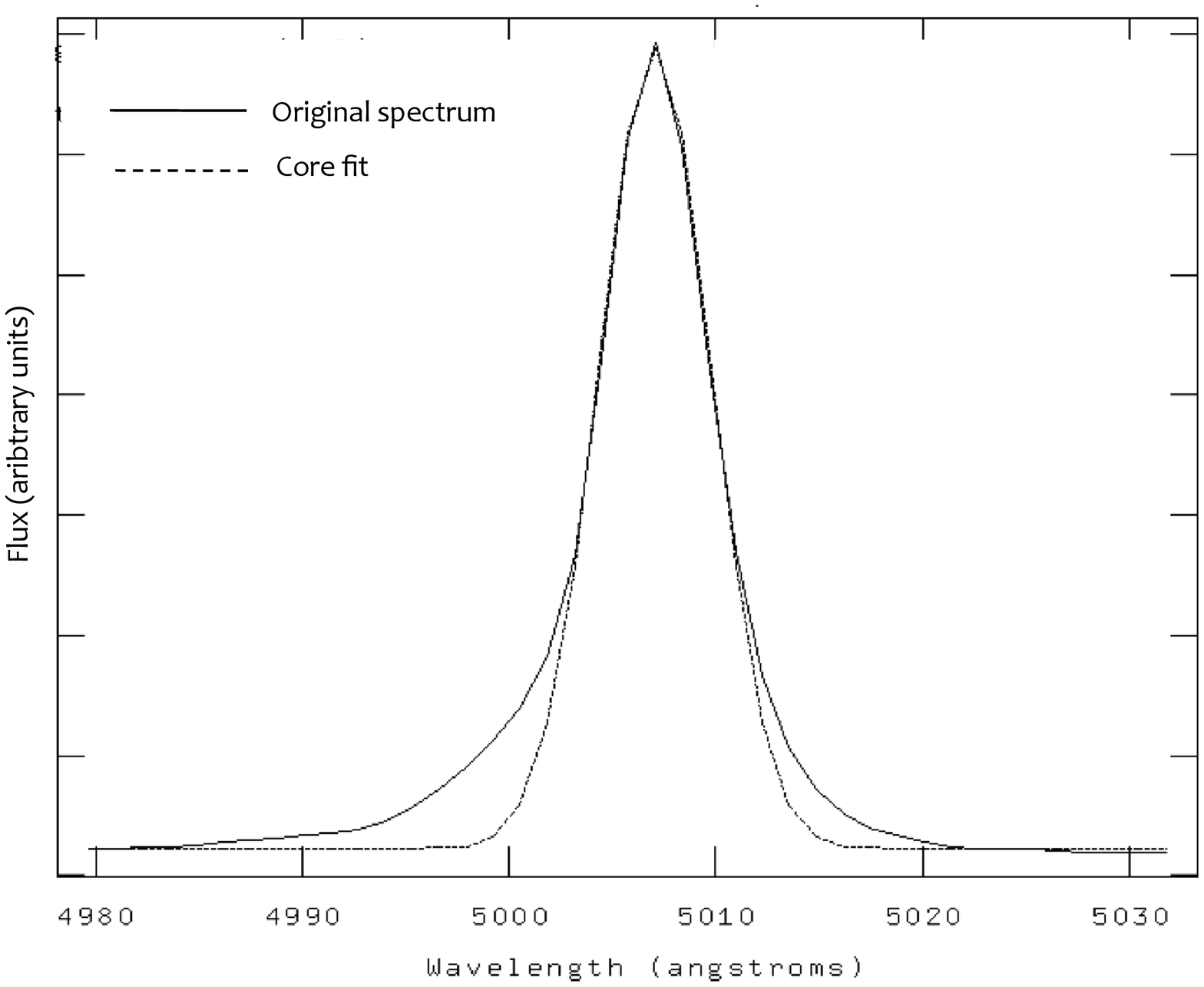}
\vspace{2.8in}
\caption{We illustrate in this figure how the narrow core of the [OIII] line is isolated. The central $\lambda$ and FWHM (corrected for instrumental
broadenning)   are used to estimate  $z_{sys}$ and FWHM$_{\star}$ for each quasar.}
\label{fig:core}
\end{figure}

No prominent stellar features are found in the QSO2 spectra. The CaII
K is detected in most objects, but  this is not likely to be a reliable tracer of the systemic 
redshift $z_{sys}$ and the velocity dispersion of the stars FWHM$_{\star}$   (Rodr\'\i guez Zahur\'\i n, Tadhunter \& Gonz\'alez Delgado \citeyear{rod09}).  As in VM11b, we have used the narrow core of the  [OIII]$\lambda$5007 line  (Fig. \ref{fig:core}) to determine    $z_{sys}$ (within $\pm$50 \kms).  FWHM$_{\star}$ was constrained using  the core FWHM    corrected for instrumental broadening\footnote{The instrumental resolution  ($\sim$175-220 \kms~ depending on the object) was measured for each quasar from the original SDSS spectra using sky lines close to the lines of interest.} and applying
FWHM[OIII]$_{core}$=(1.22$\pm$0.76)$\times$ FWHM$_{\star}$ (Greene \& Ho \citeyear{gre05}; see Table \,\ref{tab:tabfits}). 
The errors are those
produced by the mathematical fits. They are lower limits because they ignore the large scatter of the correlation.
\cite{gre05} warn about the usefulness of the FWHM[OIII]$_{core}$ vs. FWHM$_{\star}$ relationship in statistical studies, but not for individual sources.  Thus, the uncertainty in FWHM$_{\star}$ for individual objects is large. However,  as we will see, this does not affect  our analysis and conclusions. 

For each kinematic component isolated in the [OIII]  lines through the spectral fits, the flux $F$, FWHM (corrected for instrumental broadening) and velocity shift $V_S$ relative to the  narrow core were measured.
The ratios $R=\frac{\rm FWHM_{[OIII]}}{\rm FWHM_{\star}}$ (see  \cite{vm11b}), the relative contribution  to the total \hb~ flux  $R2=\frac{F(H\beta)}{F(H\beta)_{tot}}$  and the flux ratios \oiiihb, \hahb, \niiha~  and \siiha~ were computed  for each kinematic component.

The results are shown in Table \,\ref{tab:tabfits}. In general, the errors are those of  the mathematical fit. When several fits were successful within the expected constraints, 
the errors  account for the scatter of the acceptable values  as well.  The errors on  \hahb, [NII]/H$\alpha$ and [SII]/H$\alpha$ 
are  in general   lower limits, since the high number of constraints produces small mathematical errors.

 \subsection{Radio continuum}
\label{sec:data_radio}

In order to assess the radio loudness of our QSO2, we obtained their 1.4\,GHz radio flux by comparing the FIRST (Faint Images of the Radio Sky at Twenty-cm) and NVSS (NRAO VLA Sky Survey) catalogues \citep{bec95,con98}. Most sources in our sample appear unresolved at the 5'' FIRST resolution and we thus obtained the peak flux of the component that matches the QSO2 position. Only for SDSS J1229+38 and SDSS J1430+13 we took the integrated flux of the FIRST data, which was double the FIRST peak flux and matched the unresolved flux of the lower resolution NVSS data, suggesting that these two sources are marginally resolved in FIRST. SDSS J1102+64 was not covered by FIRST, hence we took the unresolved NVSS flux. SDSS J0041-09 and SDSS J1552+27 only show a tentative FIRST detection at the level of $\sim$3$\times$ the root-mean-square noise level.

In this paper we present the restframe radio power $P_{\rm 1.4\,GHz}$, which was derived by K-correcting our data assuming a power law spectral index $S_{\nu} \propto \nu^{\alpha}$, such that $P_{\rm 1.4 GHz} = 4 \pi D_{L}^{2} S_{\nu} (1+z)^{-1-\alpha}$ \citep[with $D_{L}$ the luminosity distance;][]{wri06}. We assume $\alpha = +0.094$, as found to be the median value of the spectral index between
1.4 GHz and 8.4 GHz ($\alpha_{1.4}^{8.4}$) among a sample of QSO2 by \citet{lal10}. They discuss that the $\alpha$ values for QSO2 show a bi-modal distribution between steep-spectrum radio-AGN ($\alpha_{1.4}^{8.4} \sim -0.7$) and sources with a much flatter spectrum at lower radio power ($\alpha_{1.4}^{8.4} \sim +0.6$). The K-corrections are found to be small and do not alter the conclusions from this paper.

\subsection{Far infra-red}
\label{sec:data_FIR}

IRAS 60$\mu$m detections are available for 8 of the 21 QSO2. For the remaining sources we assume an upper limit of $S_{\rm 60\mu m} \leq 0.2$\,Jy \citep{mos93}. To estimate the 60$\mu$m rest-frame fluxes and luminosities, a first-order K-correction was applied assuming a power-law spectral-energy distribution of $S_{\nu} \propto \nu^{-2}$ \citep[e.g.][]{wan10}.

To estimate the far infra-red flux $S_{\rm FIR}$ ($42.5-122.5 \mu$m) we followed \citet{hel85}:\\
\ \\
\indent \( S_{\rm FIR} = 1.26 \cdot 10^{-11}\,(2.58 \cdot S_{\rm 60 \mu m} + S_{\rm 100 \mu m}) \)\\

where $S_{\lambda}$ are the flux densities in Jy and $S_{\rm FIR}$ is in erg s$^{-1}$ cm$^{-2}$.

\noindent Because none of the QSO2 in our sample were reliably detected at 100$\mu$m, we assigned a 100$\mu$m flux
(or upper limit) assuming $S_{\rm 60\mu m}/S_{\rm 100\mu m} = 0.26$ as found in similar objects at $z \sim 0.3 - 0.4$  (Rodr\'\i guez et al. \citeyear{rod14}).

\section{Results}

\subsection{Isolating the emission from the turbulent and the ambient gas}

The results of the kinematic analysis for all objects are shown in Table \,\ref{tab:tabfits}. But for one object (SDSS J1552+27, whose   emission lines can be
fitted with a single Gaussian), the lines show complex substructure  in all other quasars with  2 or 3 kinematics components. 
In order to identify which components are turbulent compared with the underlying stellar motions we have calculated the turbulence  
parameter $R=\frac{\rm FWHM_{[OIII]}}{\rm FWHM_{\star}}$ defined by  \cite{vm11b}.

\begin{table*}
\tiny   
\centering
\begin{tabular}{lllllllllll}
\hline
(1) & (2) & (3) &  (4) & (5) & (6) &  (7) & (8)  & (9) &  (10) & (11)\\
Object & FWHM[OIII] & $V_S$ &  R  &  $F(\hb)$/10$^{-15}$ &  R2 &  \oiii/\hb &  \ha/\hb & \nii/\ha  & \sii/\ha      & FWHM$_{\star}$ \\   
 &    \kms & \kms &   & erg s$^{-1}$  cm$^{-2}$ & & &  & & &  \kms  \\ \hline
SDSS J0041-09 C1 &  305$\pm$15 & 0$\pm$9 & 1.23$\pm$0.07 & 1.7$\pm$0.2 & 0.65$\pm$0.05 & 11.3$\pm$0.8 & 4.0$\pm$0.2 & 1.08$\pm$0.06 & 0.79$\pm$0.04        &  247$\pm$9 \\ 
C2  & $\la$100 & -196$\pm$9 &  $\la$0.40 &  0.17$\pm$0.07 &  0.07$\pm$0.03 &  27$\pm$12 & 9$\pm$4 & 0.9$\pm$0.3  & 0.7$\pm$0.3  &   \\
C3 &  824$\pm$21 &  -206$\pm$15 &  3.3$\pm$0.1 &  0.73$\pm$0.14 &  0.28$\pm$0.06 &  17$\pm$3 &  3.9$\pm$0.9 &  1.7$\pm$0.4  & 1.0$\pm$0.2 & \\ \hline
SDSS J0232-08  C1 &    $\la$110 & -266$\pm$31 & $\la$0.39  & 0.4$\pm$0.04 &   0.08$\pm$0.01 	& 8.3$\pm$1.0 & 2.8$\pm$0.3 & 0.42$\pm$0.07  &  0.58$\pm$0.03 & 310$\pm$30     \\
C2	  &   	 379$\pm$31	&   0$\pm$31  & 1.2$\pm$0.2 &  2.0$\pm$0.1 &  0.38$\pm$0.02 & 	8.5$\pm$1.7 &	4.1$\pm$0.2 & 0.88$\pm$0.05 & 0.81$\pm$0.09   \\ 
C3	   &	767$\pm$39	&  -371$\pm$45  & 2.5$\pm$0.3 &  2.8$\pm$0.1 &   0.54$\pm$0.02 &  12.7$\pm$1.4  &	4.1$\pm$0.2 &  0.77$\pm$0.02   & 0.53$\pm$0.07 \\ \hline
 SDSS J0759+50 C1 &  217$\pm$7 &  -21$\pm$24 & 0.60$\pm$0.05 &   4.2$\pm$0.1 &  0.18$\pm$0.01 & 12.1$\pm$0.5 &  4.4$\pm$0.2 & 0.98$\pm$0.03   
 & 0.55$\pm$0.03 &  362$\pm$27  \\
C2	&   744$\pm$13 & 72$\pm$24  & 2.0$\pm$0.2 &   7.9$\pm$0.4  &  0.35$\pm$0.02 &   15.1$\pm$0.9  &   5.0$\pm$0.3 & 1.4$\pm$0.1 & 0.65$\pm$0.05 \\
C3	& 1709$\pm$22 & 29$\pm$27 &  4.7$\pm$0.4 & 10.7$\pm$0.5  &  0.47$\pm$0.05 &  18.3$\pm$0.9 & 5.6$\pm$0.3 & 0.82$\pm$0.05   &  0.23$\pm$0.03 \\ \hline
SDSS J0802+25     C1 &  130$\pm$13 &   -13$\pm$7 &  0.49$\pm$0.05  &  2.0$\pm$0.1 & 0.15$\pm$0.01 &   12.3$\pm$0.8 &  3.4$\pm$0.2 & 0.11$\pm$0.05 & 0.15$\pm$0.1 & 265$\pm$12 \\ 
	C2	  & 452$\pm$17	 &	 154$\pm$9  &  1.7$\pm$0.1 &   6.0$\pm$0.2 & 0.46$\pm$0.02 &       11.6$\pm$0.5   & 4.9$\pm$0.2  &  0.68$\pm$0.04 & 0.92$\pm$0.03  \\ 
	C3	    & 1358$\pm$20 	& -208$\pm$11 &   5.1$\pm$0.2 & 5.1$\pm$0.2  &  0.39$\pm$0.02 &     13.7$\pm$0.6 & 4.5$\pm$0.2 & 1.02$\pm$0.06 
		    & 0.50$\pm$0.05 & \\ \hline 
SDSS J0818+36 C1 &  334$\pm$10 & 0 $\pm$7 & 1.06$\pm$0.05 & 5.24$\pm$0.09 & 0.77$\pm$0.04 & 12.1$\pm$0.6 & 4.6$\pm$0.2 & 0.32$\pm$0.01 & 0.30$\pm$0.05 & 314$\pm$12 \\
C2 & 809$\pm$41 & -61$\pm$10 & 2.6$\pm$0.2 & 1.6$\pm$0.2 &  0.23$\pm$0.03 & 16.7$\pm$2.1 & 4.5$\pm$0.7 & 0.36$\pm$0.07 & 0.34$\pm$0.04  \\ \hline
SDSS J0936+59  C1&   497$\pm$11   &   12$\pm$9  &    1.40$\pm$0.05 &   2.70$\pm$0.09 &  0.69$\pm$0.04 &   9.3$\pm$0.6  & 3.5$\pm$0.1 &   0.73$\pm$0.03   & 0.47$\pm$0.04 &   355$\pm$11     \\
C2			 &	 1002$\pm$55 &  -80$\pm$10  & 2.8$\pm$0.2 &    1.20$\pm$0.15 &    0.31$\pm$0.04 &    9.0$\pm$1.5  &  4.8$\pm$0.7 & 0.63$\pm$0.08 & 0.44$\pm$0.09  \\ \hline
SDSS J1010+06 C1 &  423$\pm$14 & -11$\pm$9 & 0.85$\pm$0.05  &     3.8$\pm$0.1  &    0.31$\pm$0.02       & 4.9$\pm$0.2 & N/A & N/A  & N/A  &  495$\pm$26 \\
C2		 &   1163$\pm$46 & -66$\pm$13 &   2.3$\pm$0.2 &   4.1$\pm$0.3 & 0.33$\pm$0.03 &         10.1$\pm$0.8  & N/A & N/A   & N/A \\
C3		&  3414$\pm$398 &  -250$\pm$76  &  6.9$\pm$0.9  &      4.5$\pm$0.4  &  0.36$\pm$0.04        & 3.5$\pm$1.2   &  N/A & N/A  & N/A \\  \hline	
SDSS J1100+08 C1 &    333$\pm$8  & 0$\pm$7   & 0.85$\pm$0.03  &   6.6$\pm$0.2 &  0.33$\pm$0.02   &  12.4$\pm$0.5 & 3.6$\pm$0.1 & 0.74$\pm$0.01 & 0.32$\pm$0.02  &  391$\pm$11   \\
C2		 & 875$\pm$31 & -50$\pm$9  &  2.2$\pm$0.1 &  7.9$\pm$0.4 &  0.39$\pm$0.03 & 11.9$\pm$0.9 & 3.8$\pm$0.2 & 0.76$\pm$0.03 & 0.33$\pm$0.02 &   	 \\ C3	 	& 2174$\pm$138  & 15$\pm$21   & 5.6$\pm$0.4 & 5.6$\pm$0.6  &  0.28$\pm$0.02 &    11.2$\pm$1.4  & 6.7$\pm$1.1 & 0.47$\pm$0.04 & N/A &  	 \\ 	
	\hline
SDSS J1102+64  & 295$\pm$9 & 70$\pm$7 & 0.71$\pm$0.04 &	2.5$\pm$0.1 &	0.30$\pm$0.01 &  6.0$\pm$0.3  & 7$\pm$0.2  & 0.82$\pm$0.02  &  0.51$\pm$0.2 & 434$\pm$21  \\ 
C1	  & 641$\pm$9 &  -153$\pm$11 &  1.49$\pm$0.07 &   3.4$\pm$0.2 &  	0.42$\pm$0.02 &	11.8$\pm$0.9 & 6.8$\pm$0.4 & 0.90$\pm$0.04 &  0.56$\pm$0.09 \\  
C2º	   &  933$\pm$43 &  186$\pm$9 &  2.2$\pm$0.1  &   2.2$\pm$0.2 &  0.28$\pm$0.02	&	7.0$\pm$1.1 & 5.2$\pm$0.6 & 1.5$\pm$0.5 & 0.6$\pm$0.2	 \\  \hline
SDSS J1152+10 C1 &  156$\pm$18 & -309$\pm$9 & 0.70$\pm$0.09 & 3.3$\pm$0.1 &  0.26$\pm$0.01 & 13.6$\pm$0.6 & 3.4$\pm$0.1 & 0.35$\pm$0.02 & 0.25$\pm$0.05  & 223$\pm$9 \\ 
C2  & 236$\pm$13 & 12$\pm$9 & 1.06$\pm$0.07 & 6.3$\pm$0.2 & 0.51$\pm$0.02 & 13.1$\pm$0.6 & 3.3$\pm$0.1 & 0.46$\pm$0.01 & 0.44$\pm$0.01 &    \\ 
C3 &  693$\pm$32 & -88$\pm$11 & 3.1$\pm$0.2 & 2.9$\pm$0.3 & 0.23$\pm$0.02 & 11.4$\pm$1.6 & 4.6$\pm$0.5 & 0.71$\pm$0.05 & 0.48$\pm$0.04 \\ \hline
SDSS J1153+58 C1 &  240$\pm$18 & 14$\pm$8 & 0.9$\pm$0.1 &  2.7$\pm$0.1 & 0.27$\pm$0.02 & 13.1$\pm$1.3 & 3.7$\pm$0.3 & 0.79$\pm$0.08 & 0.8$\pm$0.1 & 269$\pm$23 \\
C2  &   750$\pm$24 &  -24$\pm$13  &  2.8$\pm$0.3 &  7.4$\pm$0.3 &  0.73$\pm$0.03 & 10.3$\pm$0.5 & 4.3$\pm$0.2 & 0.92$\pm$0.03 & 0.51$\pm$0.05 & \\ \hline 
SDSS J1229+38 C1 &  $\la$104 & 85$\pm$9   & $\la$0.42    &     0.30$\pm$0.04 &  0.09$\pm$0.01 &  10.5$\pm$2.2 &  2.8$\pm$0.4  & 0.7$\pm$0.1  & N/A   &  378$\pm$16 \\ 
C2 & 393$\pm$14 & -52$\pm$11 &   1.04$\pm$0.06 &      2.11$\pm$0.07 &  0.70$\pm$0.06  &  9.5$\pm$0.9   & 3.7$\pm$0.3 & 0.48$\pm$0.02  & N/A & \\
C3	 &  898$\pm$52 & -192$\pm$25   &    2.4$\pm$0.2 &  0.6$\pm$0.1   &  0.21$\pm$0.04 &  16.0$\pm$3.3   & 4.5$\pm$1.0  & 0.25$\pm$0.08  & N/A  \\ \hline 	
SDSS J1300+54 C1  & 142$\pm$12 & -3$\pm$9 & 1.22$\pm$0.15 & 9.1$\pm$0.3 & 0.71$\pm$0.03 & 12.4$\pm$1.0 & 3.4$\pm$0.1 & 0.26$\pm$0.01 & 0.28$\pm$0.01 & 116$\pm$10 \\ 
C2  & 352$\pm$23 & 54$\pm$16 & 3.0$\pm$0.3 & 3.8$\pm$0.2 & 0.29$\pm$0.02 & 11.8$\pm$2.5 & 4.1$\pm$0.3 & 0.36$\pm$0.03 & 0.30$\pm$0.02 & \\ \hline
SDSS J1405+40  C1 & 270$\pm$11 &   4$\pm$8  &    0.95$\pm$0.06 &     7.1$\pm$0.1  &   0.68$\pm$0.02 &   10.4$\pm$0.3 &  3.41$\pm$0.07 &  0.45$\pm$0.01 & 0.32$\pm$0.02 &  284$\pm$12  \\  
C2		& 830$\pm$19  &  -72$\pm$9 &   2.9$\pm$0.1   &            3.4$\pm$0.2 &  0.32$\pm$0.02 &  17.7$\pm$1.2 & 5.4$\pm$0.4  & 0.50$\pm$0.02  &  0.25$\pm$0.03 &  \\ \hline
SDSS J1430+13 C1 &   440$\pm$9 & -9$\pm$13  &   1.00$\pm$0.04 &   18.0$\pm$0.1  & 0.57$\pm$0.04 &  7.1$\pm$0.5 &  3.6$\pm$0.2 & 0.34$\pm$0.01 & 0.14$\pm$0.03  & 439$\pm$16  \\
	C2		&	1024$\pm$32 & -68$\pm$15 &  2.3$\pm$0.1 &   13.8$\pm$0.1   &  0.43$\pm$0.04 &   8.2$\pm$0.7 & 4.9$\pm$0.4 & 0.70$\pm$0.02 & 0.50$\pm$0.02  \\ \hline  
SDSS J1437+30  C1 & 154$\pm$16  & 27$\pm$9 & 0.45$\pm$0.05 & 0.8$\pm$0.2 &  0.08$\pm$0.02 & 15$\pm$3 &  3.0$\pm$0.6 & 2.3$\pm$0.2 & 1.1$\pm$0.1 & 339$\pm$11  \\ 
C2  &  497$\pm$13 & -32$\pm$9 & 1.47$\pm$0.06  & 6.7$\pm$0.02 & 0.75$\pm$0.03 &  12.6$\pm$0.3 & 4.0$\pm$0.1 & 1.25$\pm$0.02  & 0.80$\pm$0.03 &  \\
C3  & 1254$\pm$37 &  107$\pm$16 & 3.7$\pm$0.2 &  1.5$\pm$0.2 &  0.17$\pm$0.02 &  11.5$\pm$1.15 &   6.3$\pm$0.8 &  1.56$\pm$0.06 & 0.55$\pm$0.06  \\ \hline
SDSS J1440+53 C1 &   160$\pm$19   & 2$\pm$9   &  0.7$\pm$0.1  &   0.41$\pm$0.04  &    0.40$\pm$0.01 &   8.5$\pm$0.4 &   3.4$\pm$0.1 & 0.30$\pm$0.02  & 0.37$\pm$0.02 &  221$\pm$15 \\
	C2		&  536$\pm$17   & 19$\pm$11   &   2.4$\pm$0.2    &     0.38$\pm$0.01 &  0.36$\pm$0.01 &   12.8$\pm$0.4  &    4.5$\pm$0.2 &  0.42$\pm$0.03  & 0.36$\pm$0.01 & \\
 	C3		& 1836$\pm$60  & -225$\pm$27   &    8.3$\pm$0.6   &  0.25$\pm$0.01 &  0.24$\pm$0.01  &   9.2$\pm$0.6 & 4.2$\pm$0.3 & $\le$0.25 &  0.14$\pm$0.03   &   \\ \hline

SDSS J1455+32 C1 &    241$\pm$9   & 0$\pm$9    &  0.76$\pm$0.04   &   2.2$\pm$0.07  &  0.35$\pm$0.02 & 10.7$\pm$0.4 &  3.7$\pm$0.2 & 0.57$\pm$0.03  & 0.49$\pm$0.02 &  318$\pm$11  \\
		C2	&  789$\pm$15   & 106$\pm$9    &   2.5$\pm$0.1 &    3.0$\pm$0.2 &  0.48$\pm$0.04 &   15.2$\pm$1.3 &  4.9$\pm$0.3 & 0.67$\pm$0.03  &  0.35$\pm$0.01  \\
 	C3	&  1253$\pm$68  & -116$\pm$46   &  3.9$\pm$0.3  &    1.1$\pm$0.2  &  0.17$\pm$0.03 &   13.0$\pm$3.5 & N/A  & N/A & N/A \\ \hline

SDSS J1552+27 & 272$\pm$13 & 9$\pm$9 & 1.2$\pm$0.1 & 4.9$\pm$0.1 & 1.0 & 13.7$\pm$0.3 & 3.9$\pm$0.08 & 0.60$\pm$0.01 & 0.43$\pm$0.01 & 233$\pm$15 \\ \hline
SDSS J1653+23  C1 &   170$\pm$10   & 56$\pm$13 &  0.65$\pm$0.05     &    3.1$\pm$0.1 &   0.27$\pm$0.01 &   16.8$\pm$0.8  & 3.0$\pm$0.2  & 0.43$\pm$0.03  &  0.32$\pm$0.08 &  260$\pm$11      \\
C2		&		405$\pm$16   &  -55$\pm$15  &   1.56$\pm$0.09 &  6.8$\pm$0.2 & 0.60$\pm$0.03  &   8.8$\pm$0.4 &  4.0$\pm$0.2 &  0.31$\pm$0.02 & 0.36$\pm$0.04 &  \\
C3		&	 	969$\pm$61 	&  -87$\pm$14  &  3.7$\pm$0.3    &    1.5$\pm$0.2  &  0.13$\pm$0.02  &  14.8$\pm$2.5  &  5.9$\pm$1.0 &  0.47$\pm$0.04 & 0.34$\pm$0.07 \\ \hline
SDSS J2134-07 C1 &  345$\pm$10 &  0$\pm$9 & 1.22$\pm$0.06 & 2.7$\pm$0.1 & 0.82$\pm$0.05 & 13.0$\pm$0.7 & 3.9$\pm$0.2 & 0.50$\pm$0.01 & 0.45$\pm$0.01 & 283$\pm$13 \\
C2 &  859$\pm$90 & 60$\pm$20 & 3.0$\pm$0.4 & 5.9$\pm$1.5 & 0.18$\pm$0.05 & 14$\pm$4 &  4.9$\pm$1.3 & 0.45$\pm$0.07 & 0.37$\pm$0.03 \\  \hline
\end{tabular}
\caption{Measured parameters for the kinematic components isolated in all QSO2. $V_S$ (3) is the velocity shift relative to the narrow core of [OIII]$\lambda$5007. $R$=FHWM[OIII]/FWHM$_{\star}$  (4)  parametrizes the level of turbulence of the gas relative to the stars. $R2$=$\frac{F(H\beta)}{F(H\beta)_{tot}}$  (6) 
is the ratio of the \hb~ flux of a given kinematic component  (5) relative to the total \hb~flux.  FWHM$_{\star}$=FWHM[OIII]$_{core}$/1.22 (column 11), following  Greene \& Ho (2005) (see \S3.1).
C1, C2, C3 refer to different components for a given object. N/A means that
no satisfactory fit could be obtained for those lines.}  
\label{tab:tabfits}
\end{table*}

\begin{figure*}
\includegraphics{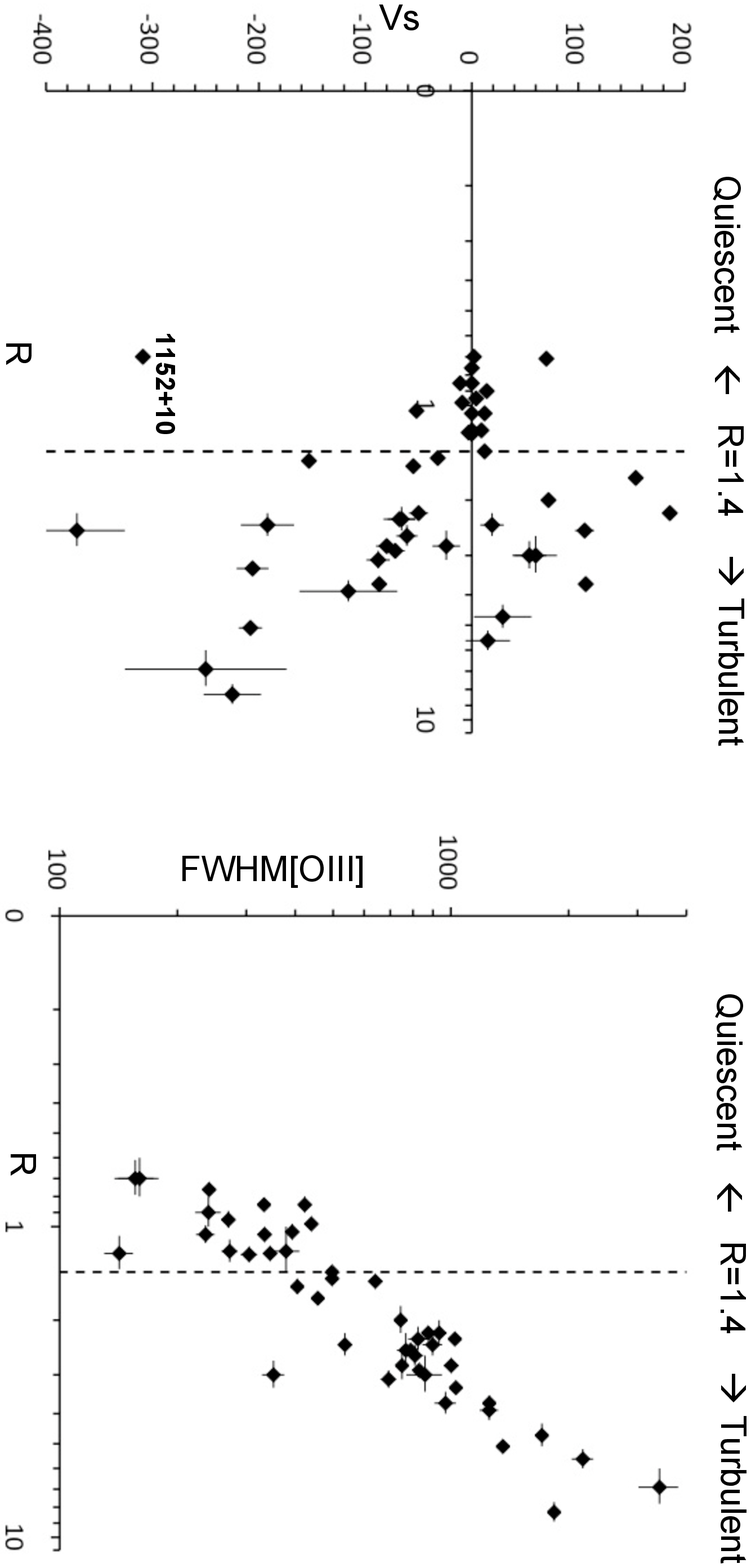}
\vspace{3in}
\caption{Identifying turbulent gas  in the nuclear region of QSO2 at $z\sim$0.1. For each kinematic component, the
velocity shift  $V_S$ relative to the narrow [OIII] core  and the FWHM are plotted vs. $R$. At $R>$1.4, a sudden change in the behavior of $V_S$ is appreciated. A trend for larger $|V_S|$ values and a preference for blue shifts appear. We propose that this value marks the approximate transition between the turbulent and quiescent regimes. SDSS 1152+10 shows double peaked emission lines and the determination of $z_{sys}$ is uncertain.}
\label{fig:beh}
\end{figure*}

We identify 3 different kinematic regimes, according to the $R$ value.  Fig. \ref{fig:beh}, where we plot $V_S$ and FWHM vs. $R$ for all  kinematic components
with $R>$0.7, will help to illustrate this.

\begin{itemize}

\item {\it $R<0.7$ regime}. These components have very narrow FWHM compared with the stars. They  are 
found  in 7/21 QSO2 (Table \,\ref{tab:tabfits}). They are always relatively faint, with a contribution to the total H$\beta$ flux $<$20\% in general ($R2<$0.2). Its physical nature is uncertain but  it could be related to the 
presence within the 3" fiber of companion emission line nuclei, compact star forming knots,  tidal tails or other gaseous features which do not necessarily follow the stellar
kinematics of the quasar host galaxy. Such spatial components emitting very narrow lines have been identified in morphological and spectroscopic studies of QSO2  at $z\sim$0.3-0.4 
(e.g. Villar Mart\'\i n et al. \citeyear{vm11a}).

Since they are not relevant to the main purpose of this paper, we will ignore the $R<$0.7 components in the analysis that follows.

\item {\it Quiescent regime: 0.7$\le R \le$1.4}. These components have FWHM$\sim$150-500 \kms~ (Fig. \ref{fig:beh},  right) and are cluttered around $V_S\sim$0$\pm$20 \kms~ (Fig. \ref{fig:beh}, left).  Both $R$  and the 
small $V_S$ values are consistent with the stellar motions. This is expected since they often fit approximately the narrow core of the line which
has been used to calculate the stellar velocity dispersion and the systemic velocity. SDSS J1152+10 in an exception. It shows a kinematic component with $V_S\sim$-300  \kms~  and small $R\sim$0.7.  However, the spectrum of this object shows double peaked emission lines and the determination of $z_{sys}$ is uncertain. 

\item  {\it Turbulent regime: $R>$1.4}. This $R$ value marks a sudden transition on the behavior of the kinematic components.
A large scatter  appears for $V_S$, which is in the range $\sim$[-370,+200] \kms~(Fig. \ref{fig:beh}, left).  The FWHM spans a range $\sim$350-3400 \kms~  but it is $\ga$600 \kms~ in most cases (Fig. \ref{fig:beh}, right). 
All QSO2 in the sample but the quiescent SDSS J1552+27 (20/21) have at least one kinematic component with $R>$2. Even accounting for the large scatter of the FWHW[OIII] vs. FWHM$_{\star}$ relation (see \S3.1), such high $R$ values  indicate turbulence.  
This is further reinforced by the frequent large values and scatter of $|V_S|$ compared with the quiescent regime.
The turbulent components show a preference for    blueshifts, specially at the largest $R$ (Fig. \ref{fig:beh}, left). 

The average $R$, FWHM[OIII], $|V_S|$ and $V_S$  are shown for the turbulent and quiescent components in Table \,\ref{tab:median}.

Based on these results, we identify the turbulent components  with emission from outflowing gas (see also \cite{vm11b}). 
Differential reddening explains the more frequent blueshifts, so that the receding (redshifted) gas is more obscured than the approaching (blueshifted) gas.  
Only 4 turbulent components have positive $V_S>$+50 \kms~ ( taking errors
into accoun)t vs. $\sim$13 with confirmed $V_S<$-50 \kms~ (i.e. most probably larger than the uncertainty relative to $z_{sym}$, see \S3.1).  The  4 objects where the redshifted turbulent gas has been detected 
(SDSS J0759+50, SDSS J0802+25, SDSS J1102+64, SDSS J1455+32) show  {\it moreover} a turbulent  component blueshifted by a similar amount. In these objects, we are probably observing both the receding and the approaching sides  of the expanding outflow.

Thus, we confirm the presence of nuclear ionized outflows in 20 out of the 21 QSO2 at $z\la$0.1, which are responsible for the turbulent motions identified
in these objects.
On the other hand, we identify the quiescent components with ambient gas  which apparently has not been reached by the outflows. 
Based on the analysis above, we define $R$=1.4 as the dividing  line between the quiescent and the turbulent regimes.

\begin{figure*}
\includegraphics{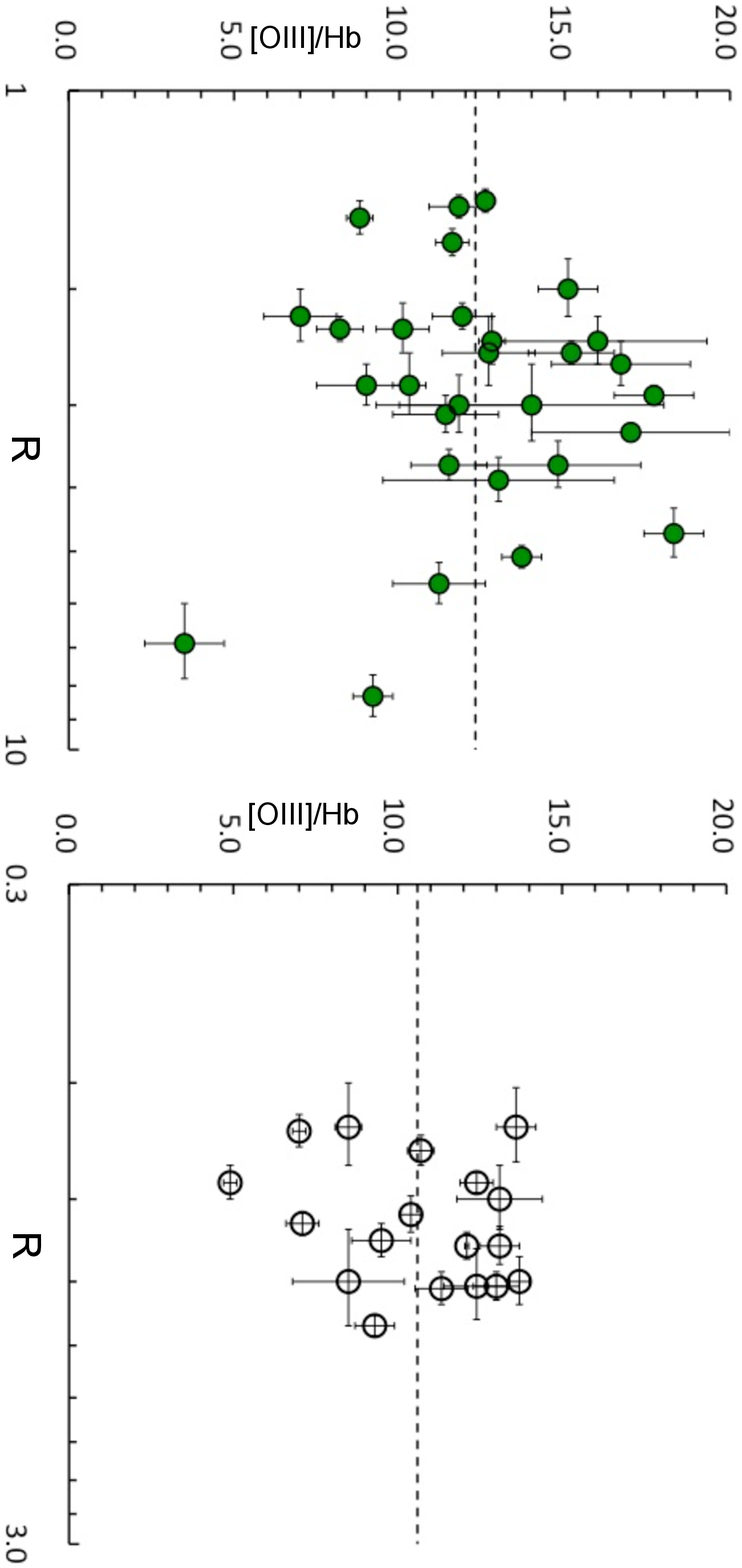}
\vspace{3in}
\includegraphics{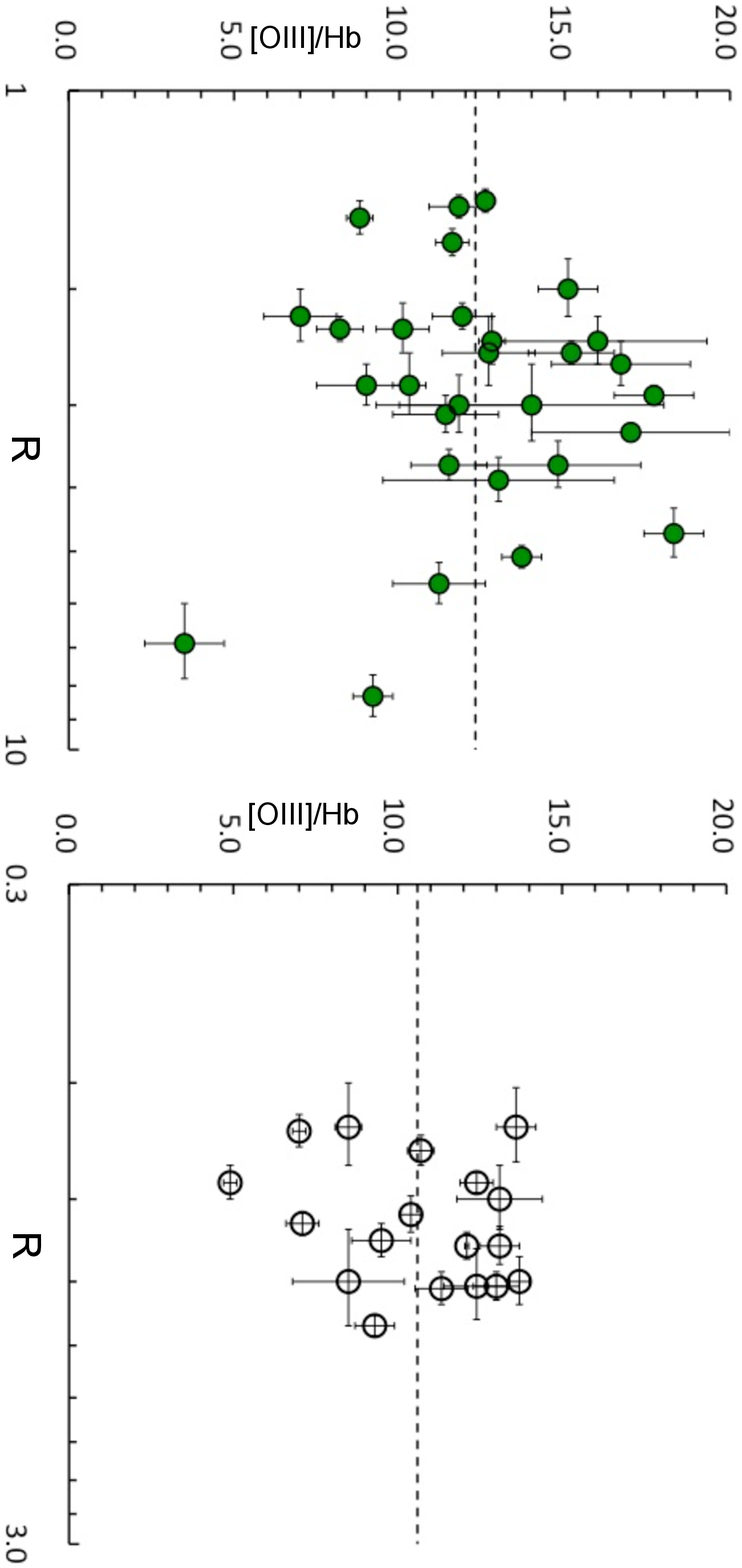}
\vspace{3in}
\includegraphics{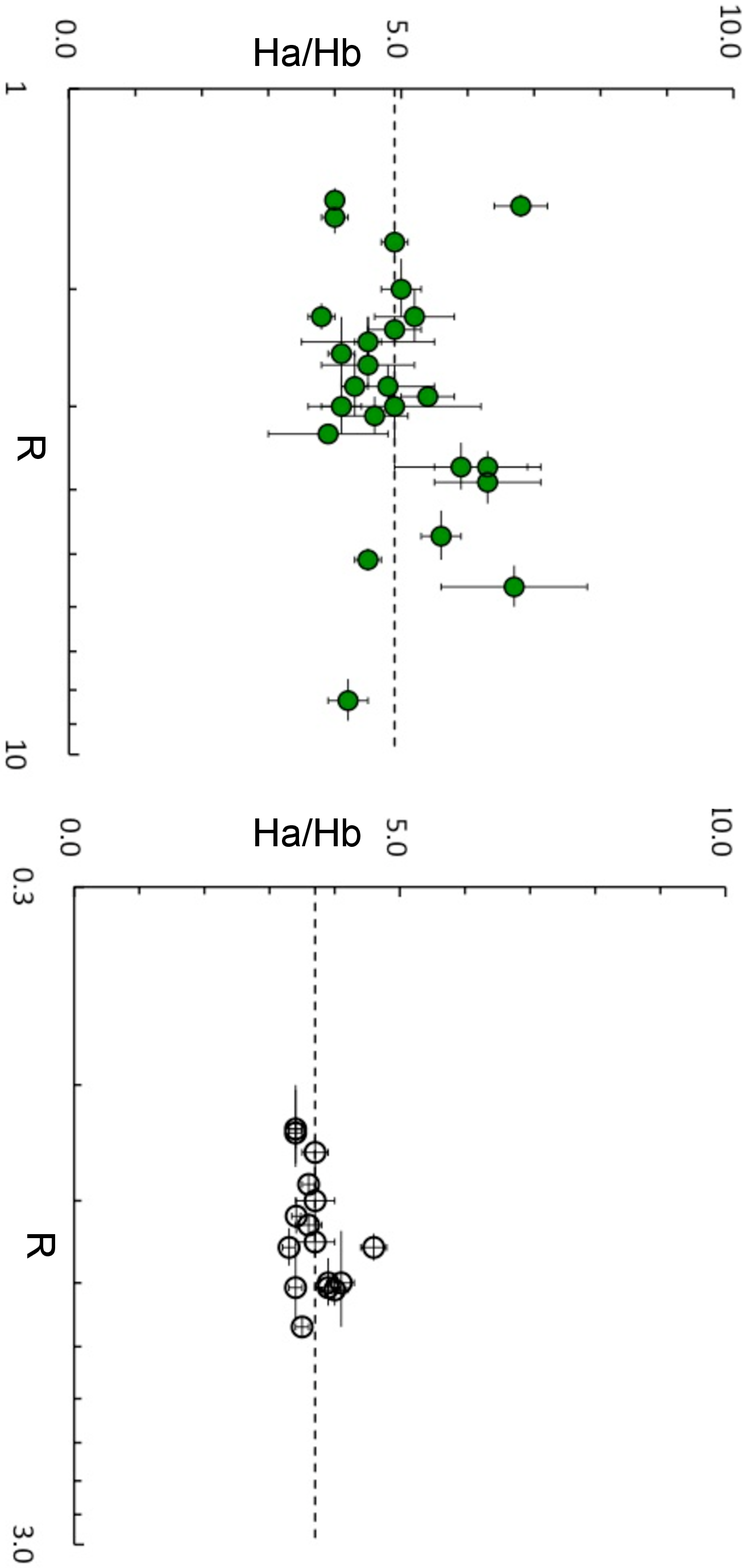}
\vspace{2.9in}
\caption{Comparison of $R2$=$\frac{F(H\beta)}{F(H\beta)_{tot}}$  (top), [OIII]/H$\beta$ (middle) and H$\alpha$/H$\beta$ (bottom) vs.  $R$ for the turbulent (left) and quiescent 
(right) components.  The average values are shown as horizontal dashed lines. The 5 objects separated by an inclined dashed line 
in the top left panel stand out above the general trend
defined by the turbulent gas. As we will see in \S4.3, the radio structures are likely to play a major role in driving the outflows in these QSO2.}
\label{fig:comp}
\end{figure*}

\end{itemize}

\subsection{Ambient vs. outflowing gas: properties comparison}
  \begin{table*}
\centering
\begin{tabular}{llllllllll} 
\hline
Regime &  $R$ & FWHM[OIII] &  $V_S$ & $|V_S|$  & $R2$ & [OIII]/H$\beta$ & H$\alpha$/H$\beta$ & \niiha & \siiha  \\ 
	&	&	\kms & \kms & \kms & & & & & \\ \hline
Quiescent & 0.98$\pm$0.01 &  287$\pm$7 & -15$\pm$4 & 28$\pm$4 & 0.55$\pm$0.01 & 10.6$\pm$0.1 & 3.68$\pm$0.02 & 0.61$\pm$0.02 & 0.44$\pm$0.01 \\
Turbulent & 3.19$\pm$0.06 & 1035$\pm$22 & -55$\pm$4  &  111$\pm$3 &  0.36$\pm$0.05 & 12.3$\pm$0.1 & 4.91$\pm$0.03 & 0.77$\pm$0.02 & 0.41$\pm$0.01  \\
\hline
\end{tabular}
\caption{Comparison of the main properties of the quiescent and turbulent components. The mean and standard error of the mean values are quoted.
 $|V_S|$ is calculated using the modulus of $V_S$ for all kinematic components.} 
\label{tab:median}
\end{table*}

By comparing the behavior and properties of the turbulent (outflowing) and quiescent (ambient) kinematic
components, we  will investigate next the origin of the outflows and their impact on the physical, kinematic and ionization properties of the 
ambient gas.

We plot in Fig. \ref{fig:comp}  $R2$=$\frac{F(H\beta)}{F(H\beta)_{tot}}$, \oiiihb~and \hahb~ vs. $R$ for all kinematics components in the quiescent (right) and turbulent (left) regimes. The average values are shown in every plot as horizontal dashed lines (see also Table \,\ref{tab:median}).

Some clear trends appear:

\begin{itemize}

\item  $R2$ vs. $R$ (Fig. \ref{fig:comp}, top). $R2$ decreases with increasing $R$ for the outflowing gas.  A Spearman's correlation analysis results on an (anti)correlation coefficient
$r=$-0.47. A Student's $t-$test gives  $t$=2.8 and  a statistically significance level of 99.0\%. 
The opposite trend is found for the ambient gas ($r$=0.71, $t$=4.0 and 99.8\% significance level).
Thus, the relative contribution of the ionized outflow 
to the total \hb ~flux decreases as the turbulence increases (although with a large scatter, the same behavior is found for the \oiii~line).
The same result was found by VM11b in their sample of  SDSS QSO2 at $z\sim$0.3-0.4. In fact, their objects overlap
completely with our sample in this diagram.

 \begin{figure*}
\includegraphics{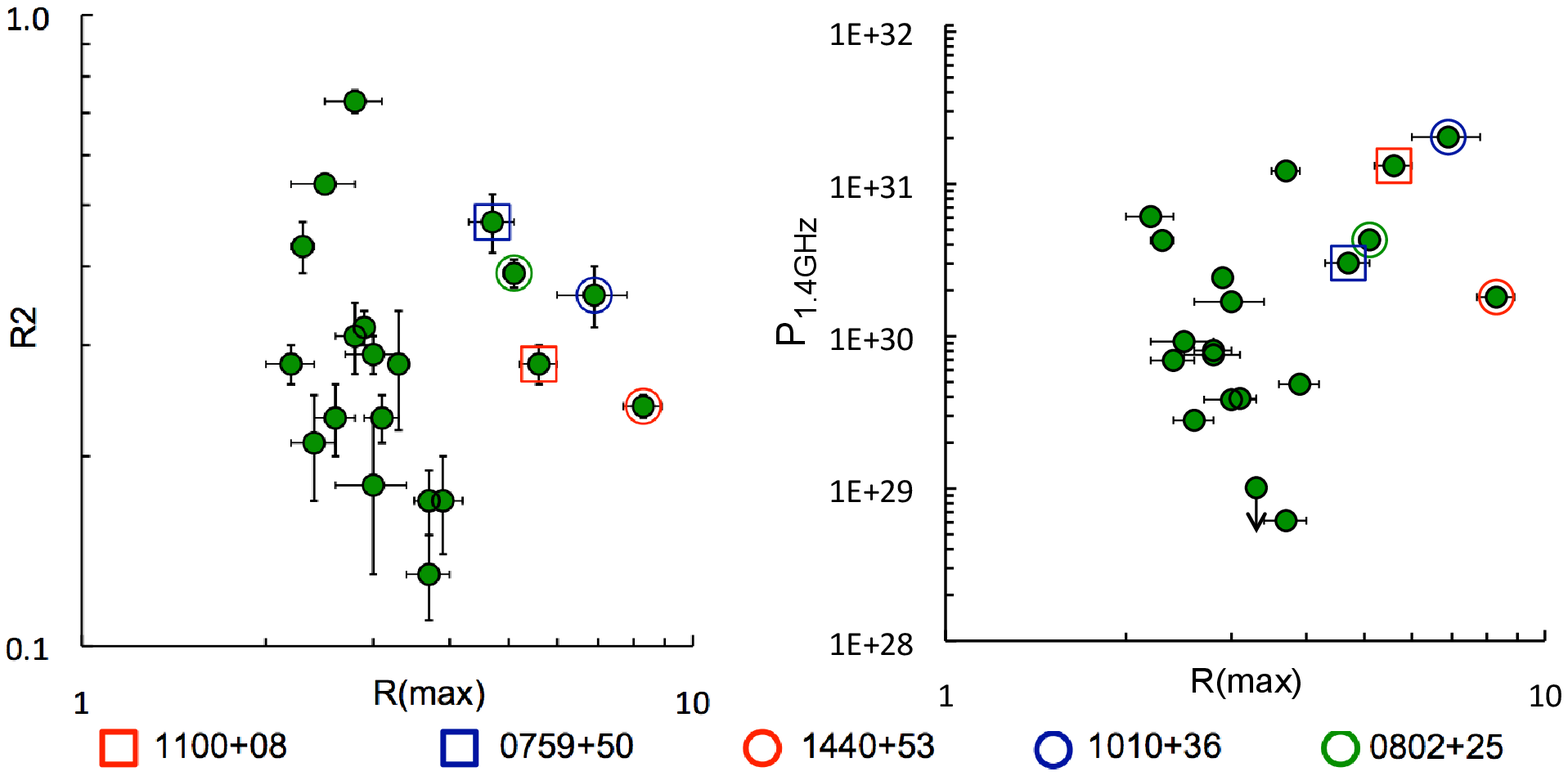}
\vspace{3.1in}
\caption{$R2$ {\sl (left)} and $P_{\rm 1.4\,GHz}$ {\sl (right)} are plotted vs. $R_{max}$. The objects highlighted in color  (Group 1) show the most
extreme turbulence ($R\ge$4.7). They are among the most radio luminous objects in the sample. The rest of the objects form Group 2 (see text).}
\label{fig:R_radio}
\end{figure*}

There are five objects that stand out above the general trend defined by the turbulent components (Fig. \ref{fig:comp}, top left panel): SDSS J0759+50, 
 SDSS J0802+25, SDSS J1010+06, SDSS J1100+08 and SDSS J1440+53. These objects will be discussed in detail in later sections.
They are of particular interest because of the extreme
turbulence  ($R\ge$4.7) and the different behavior they show. If these  objects are excluded from the statistical analysis, we obtain $r=$-0.69, $t$=4.4 and a significance level of 99.9\%.

\item ~[OIII]/H$\beta$  vs. $R$ (Fig. \ref{fig:comp}, middle).  No correlation  is found, although a trend of increasing \oiiihb~with $R$ is hinted for the turbulent components (if the 5 outliers described above are excluded, $r=$0.38, $t=$1.7 and a 90\% significance level are derived). 

The turbulent and quiescent components show  mean  [OIII]/H$\beta$=12.3$\pm$0.1
and 10.6$\pm$0.1 respectively (Table \,\ref{tab:median}),  implying higher values for the outflowing gas. Indeed, when looking at individual objects, in most cases it is found that the  most turbulent gas
 has similar or higher \oiiihb~ than the quiescent gas (Table \,\ref{tab:tabfits}).  This is also consistent  with \cite{vm11b} (see also Veilleux \citeyear{vei91}).   Extreme  \oiiihb$\ge$15 values are only shown by the outflowing gas in some objects (Fig. \ref{fig:comp}).

\item H$\alpha$/H$\beta$ vs. $R$ (Fig. \ref{fig:comp}, bottom). The outflowing gas shows higher reddening  (\hahb) than the ambient gas.   The mean values (Table \,\ref{tab:median})
are  4.91$\pm$0.03 and 3.68$\pm$0.02 for the turbulent and quiescent components respectively.

\end{itemize}

\subsection{The origin of the radio emission and its relation with the outflows}

Figure \ref{fig:R_radio} (left) shows again $R2$ vs. $R$, but   now we only take into account  for each object the broadest kinematic component,  $R_{max}$. The five objects with the largest $R_{max}$ ($\ge$4.7) appear to lie above the general trend, defining their own anticorrelation, as explained in \S4.2. 
From now on we will refer to these QSO2 as Group 1, and the rest of the objects ($R_{max}=$2-3) as Group 2.

The QSO2 in Group 1 are among the objects with the highest 1.4\,GHz radio power in our sample (Fig. \ref{fig:R_radio}, right). These plots suggest that the same source responsible for the radio emission is responsible
for inducing at least the most extreme outflows. Thus, it is of particular interest to understand the nature of the radio emission.

\begin{figure}
\includegraphics{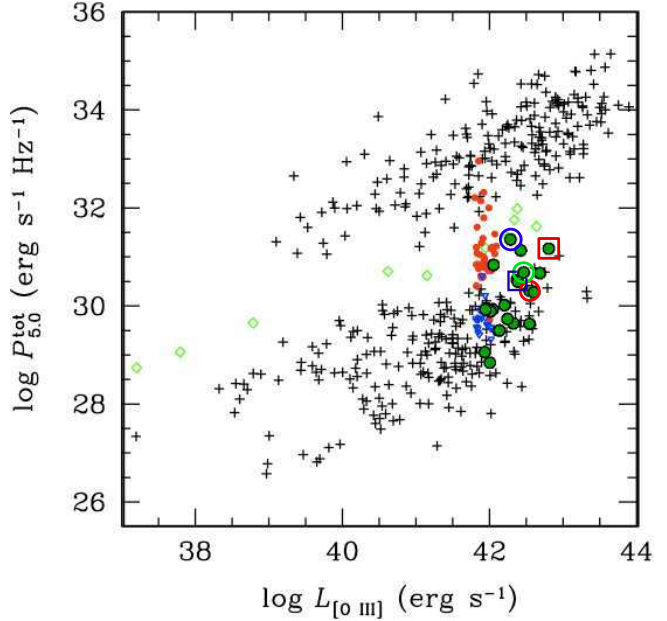}
\vspace{3.2in}
\caption{Rest-frame 5\,GHz radio power plotted against \oiii\ luminosity, comparing our QSO2 (green circles) with the QSO2 sample by \citet{lal10} at $z\sim$0.3-0.65 (red dots for detections and blue triangles for P$_{\rm 5\,GHz}$ upper limits). The crosses show a bi-modal radio-loud/-quiet distribution of a heterogeneous sample of AGN compiled by \citet{xu99}. The green diamonds are their radio intermediate objects. The QSO2 in our sample in Groups 1 and 2  are represented  as in Fig.~7. }
\label{fig:lalho}
\end{figure}

\begin{table*}
\centering
\begin{tabular}{lccccc} 
\hline
Object         &  $R_{max}$      &  $R2$         &  log(P$_{\rm 1.4\,GHz}$)  &  $q$      \\ 
               &                 &               &   (erg\,s$^{-1}$\,Hz$^{-1}$) & \\ \hline
SDSS J0041-09  & 3.3$\pm$0.8	& 0.28$\pm$0.06 & $\le$29.0 &  -     \\
SDSS J0232-08  & 2.5$\pm$0.3	& 0.54$\pm$0.02 & 30.0 &  $\le$2.1       \\ 
{\bf SDSS J0759+50}  & 4.7$\pm$0.4	& 0.47$\pm$0.05 & 30.5 & 1.6                 \\
{\bf SDSS J0802+25}  & 5.1$\pm$0.2	& 0.39$\pm$0.02 & 30.6 &  1.6                \\
SDSS J0818+36  & 2.6$\pm$0.2	& 0.23$\pm$0.03 & 29.4 &  $\le$2.4   \\
SDSS J0936+59  & 2.8$\pm$0.1	& 0.31$\pm$0.04 & 29.9 & $\le$2.1     \\
{\bf SDSS J1010+36}  & 6.9$\pm$0.9	& 0.36$\pm$0.04 & 31.3 & 1.2                \\
{\bf SDSS J1100+08}  & 5.6$\pm$0.4	& 0.28$\pm$0.02 & 31.1 & $\le$0.95        \\
SDSS J1102+64  & 2.2$\pm$0.2	& 0.28$\pm$0.02 & 30.8  & 1.7                \\
SDSS J1152+10  & 3.1$\pm$0.2    & 0.23$\pm$0.02 & 29.6  & $\le$2.1   \\
SDSS J1153+58  & 2.8$\pm$0.3	& 0.73$\pm$0.03 & 29.9 & $\le$1.8       \\
SDSS J1229+38  & 2.4$\pm$0.2	& 0.21$\pm$0.04 & 29.8  & $\le$2.3      \\
SDSS J1300+54  & 3.0$\pm$0.3	& 0.29$\pm$0.02 & 29.6 &  $\le$2.4    \\
SDSS J1405+40  & 2.9$\pm$0.1	& 0.32$\pm$0.02 & 30.4 & 1.6                 \\
SDSS J1430+13  & 2.3$\pm$0.1	& 0.43$\pm$0.04 & 30.6  & 1.4              \\
SDSS J1437+30  & 3.7$\pm$0.2	& 0.17$\pm$0.02 & 31.1  & 0.98              \\
{\bf SDSS J1440+53}  & 8.3$\pm$0.6	& 0.24$\pm$0.01 & 30.3  & $\le$0.91     \\
SDSS J1455+32  & 3.9$\pm$0.3	& 0.17$\pm$0.03 & 29.7  & $\le$2.3     \\
SDSS J1552+27  & -       	& -             & $\le$28.8 & -      \\
SDSS J1653+23  & 3.7$\pm$0.3	& 0.13$\pm$0.02 & 30.2 & 2.3               \\
SDSS J2134-07  & 3.0$\pm$0.4	& 0.18$\pm$0.05 & 29.9  & $\le$2.1   \\ \hline
\end{tabular}
\caption{Data used in the investigation of the possible link between the radio luminosities and
radio loudness of the QSO2 with the maximum gas turbulence (Fig.\,\ref{fig:R_radio} and Fig.\,\ref{fig:FIRradio}). The QSO2 in Group 1 are highlighted in bold. Notice that SDSS J1552+27 shows no evidence for turbulent gas.} 
\label{tab:rmax-q}
\end{table*}

At the radio luminosities of our sample QSO2 ($28.8 \leq {\rm log}{\rm \left(\frac{P_{\rm 1,4\,GHz}}{erg\,s^{-1}\,Hz^{-1}}\right)} \leq 31.3$, Table \,\ref{tab:rmax-q}), the radio emission can originate from star formation  or from AGN activity (non-thermal synchrotron radiation from a radio source).

In Fig.\,\ref{fig:lalho} we compare the radio power vs. \oiii\ luminosity of our quasars with that of  a heterogeneous sample of AGN, which shows a bi-modal distribution between radio-loud and radio-quiet objects (see \citealt{xu99} and \citealt{lal10} for details).  The P$_{\rm 5\,GHz}$ values for our QSO2 have been extrapolated from the K-corrected rest-frame P$_{\rm 1.4\,GHz}$ values (Sect. 3.2 and Table 4) assuming the median spectral index of $\alpha_{\rm 1.4\,GHz}^{\rm 5\,GHz} = +0.094$ ($S_{\nu} \propto \nu^{\alpha}$) found by \citet{lal10}.  According to this plot, most our QSO2 are radio quiet, except the most radio luminous objects  which lie in the intermediate region between the radio quiet and radio loud AGN.
The 5 QSO2  in Group 1   are highlighted in color. Their high radio luminosity relative to the rest of the sample is again obvious.

\begin{figure*}
\includegraphics{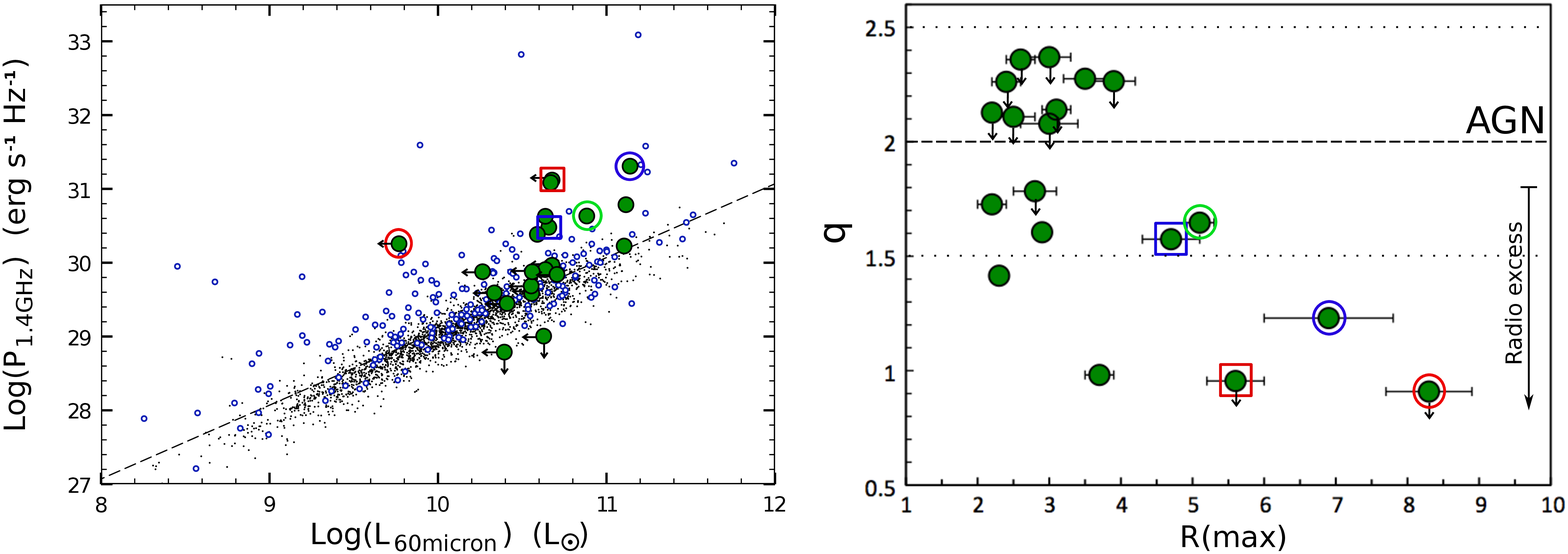}
\vspace{2.5in}
\caption{{\sl Left:} radio-FIR correlation based on the 1.4\,GHz radio continuum and 60$\mu$m IRAS luminosity, as discussed by \citet{yun01}. The QSO2 (green circles) are overplotted onto 2690 starforming galaxies (dots) and 208 radio-loud AGN (blue circles) from the 6dFGS-NVSS-FSC sample by \citet{mau07}. The dotted line corresponds to the linear radio-FIR correlation found for FIR-selected galaxies ($S_{\rm 60\mu m} \geq 2$\,Jy) by \citet{yun01}. The QSO2 that are highlighted are the same as in Fig. \ref{fig:comp} (top left panel). {\sl Right:} `$q$'-parameter (logarithmic FIR/radio flux-density ratio, see text for details) plotted against the maximum turbulence $R_{max}$ for our QSO2 (same symbols as in Fig. ~7). The average q-value (and rms scatter) found by \citet{mau07} for their AGN is shown by the dashed (dotted) line. At $q < 1.8$, the radio power is more than three times the mean value for starforming galaxies. The two QSO2 in our sample with a tentative radio detection and an IRAS 60$\mu$m upper limit are omitted from this plot.}
\label{fig:FIRradio}
\end{figure*}

What is the nature of the  radio continuum? To investigate this issue, we compare in Fig.\,\ref{fig:FIRradio} (left) our sample of QSO2 with the  `radio-FIR' correlation between the rest-frame 1.4\,GHz radio power and the FIR luminosity found for star-forming galaxies \citep[e.g.][]{kru73,con92,yun01,mau07,mor10} and AGN  \citep{mau07}.  $L_{\rm 60\mu m}$ is the rest frame luminosity value  \citep[as defined by][]{yun01}\footnote{K-corrections to P$_{\rm 1.4\,GHz}$ for the star-forming galaxies and AGN have been applied as per \citet{mau07}, while corrections to our QSO2 and the $L_{\rm 60\mu m}$ values are described in Sect.\,\ref{sec:data_FIR} $\&$ \ref{sec:data_radio}.} Although  AGN follow the correlation, many have a radio excess that causes them to lie above the  star-forming galaxies. 

Fig.\,\ref{fig:FIRradio} shows that most our QSO2 follow the general trend of AGN, while the five objects with the largest $R_{max}$ (highlighted in colour) have a radio power that is significantly larger than that expected to originate from star formation. 

A more quantitative approach can be taken by looking at the $q$ parameter defined by \citet{hel85} as:\\
\ \\
\indent \(q = {\rm log}\left([S_{\rm FIR}/3.75\times10^{12}\,{\rm Hz}]/S_{\rm 1.4\,GHz}\right)\)\\
\ \\
\noindent Therefore, $q$ is a measure of the FIR/radio flux-density ratio. The derivation of the FIR fluxes (and upper limits) for our QSO2 has been discussed in Sect.\,\ref{sec:data_FIR}. The star forming galaxies studied by \citet{mau07} show an average of $q_{\rm SF}$\,=\,2.3 with rms scatter of $\sigma_{\rm SF}$=0.18 (see also \citealt{con02}), while the AGN by \citet{mau07} show $q_{\rm AGN}$\,=\,2.0 with $\sigma_{\rm AGN}$=0.5. For sources with $q$\,$\leq$\,1.8, the radio emission is at least 3$\times$ that expected from star formation based on the radio-FIR correlation. 

A main result from our work is that Fig.\,\ref{fig:FIRradio} (right) clearly shows that   the most extreme turbulence ($R_{max}\ge$4.7, QSO2 in 
Group 1) is found at the smallest $q$ values. These QSO2 also show high radio luminosities. Thus, we propose that they host a radio source
which  plays a crucial role in driving the most turbulent gas outflows.
Objects with $R_{max}=$2-3 (Group 2)  show  no obvious link between $R_{max}$
and $q$.  On average,  they show no or small radio excess and lower radio luminosities  than Group 1. This suggests a different triggering mechanism of the outflow, unrelated to the radio source (\S5). Interestingly, the only QSO2 with no detected ionized outflow (SDSS J1552+27)
has the lowest radio luminosity, further reinforcing the link between the most extreme outflows and the radio structures.

%This, combined with the fact that the most radio-luminous QSO2 also show the most extreme and complex line profiles (large $R_{max}$ and $V_S$), strongly suggests that radio sources play a crucial role in driving the most turbulent gas outflows in our sample of optically selected QSO2.

\subsubsection{No evidence for neutral winds}
\label{sec:neutral}

%The SDSS spectra of our sample sources also cover the NaID$\lambda\lambda$5890,5896 absorption feature, allowing a search for neutral gas in the interstellar medium (ISM) of our QSO2. 

\begin{figure}
\centering
\includegraphics{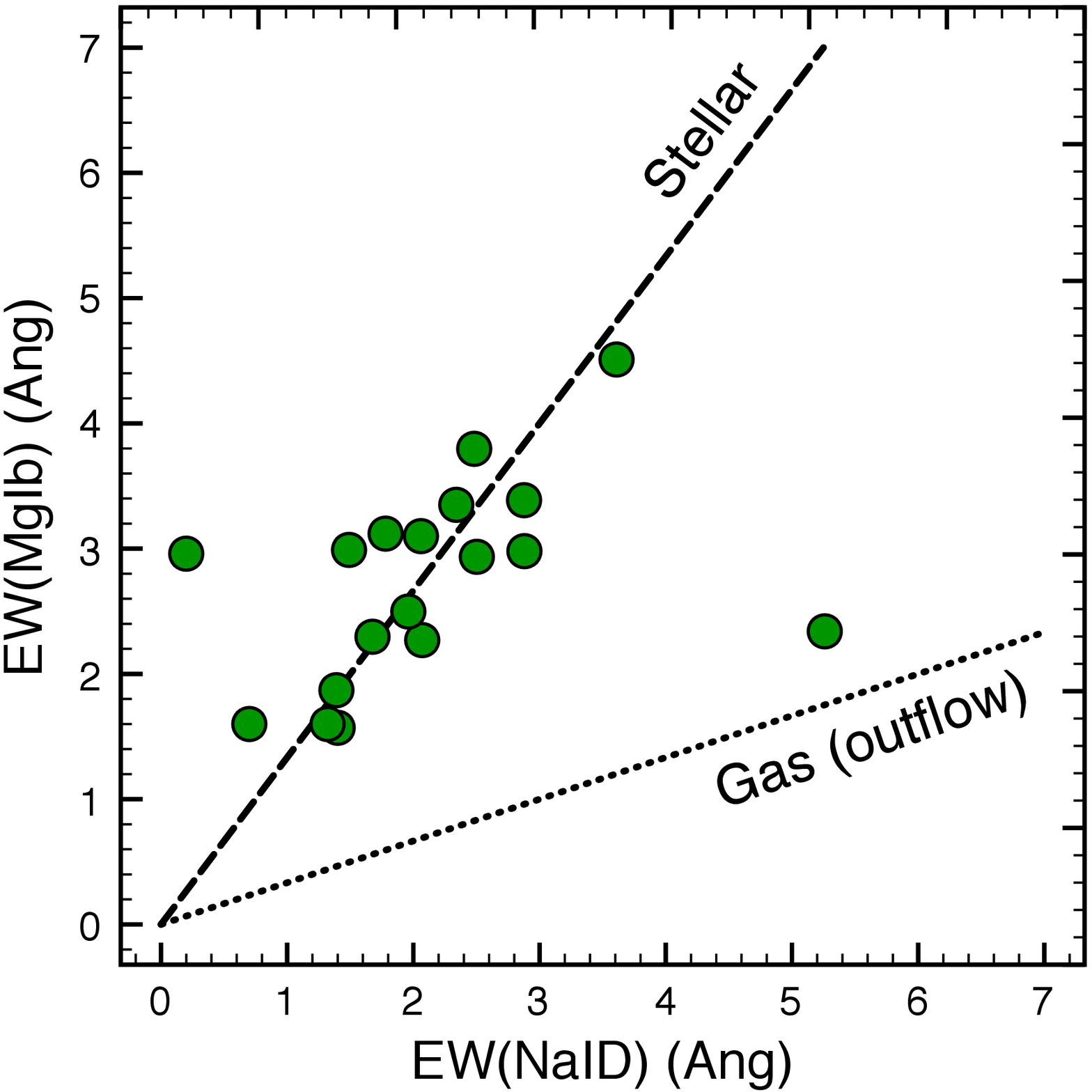}
\vspace{3in}
\caption{Equivalent width (EW) of the NaID vs. MgIb absorption features of our QSO2. The dashed line indicates the ratio at which the NaID is expected to be predominantly stellar \citep{hec00}. The dotted line corresponds to the ratio where \citet{rup05} find neutral winds due to a prominent inter-stellar NaID gas component. The source with a NaID excess due to neutral gas is SDSS\,J1102+64, which is also among the brightest FIR emitters in our sample.}
\label{fig:EW}
\end{figure}

The NaID$\lambda\lambda$5890,5896 absorption feature in the spectra of galaxies can be attributed to stellar absorption, or absorption by neutral gas in the ISM. A simple diagnostic to estimate whether there is a significant contribution of neutral ISM is by comparing the equivalent width (EW) of NaID with that of the purely stellar absorption triplet of MgIb(5167,5173,5184\,\AA) \citep[e.g.][]{hec00,mar05,rup05,kac11}. This is shown in Fig.\,\ref{fig:EW}. The EW of NaID was determined after correcting for emission-line infilling by He\,I(5876\,\AA) (which FWHM was constrained to be that of H$\beta$), while for MgIb a first order correction for potential contamination by various emission lines was applied.

The upper dashed line represents the expected stellar contribution to EW$_{NaID}$, EW$^{\star}_{NaID}$ = 0.75$\times$EW$_{MgIb}$ (Heckman et al. \citeyear{hec00}). Deviations
from this line are due to the presence of interstellar absorption in the NaID feature. 
 The lower line in this plot corresponds to EW$^{\star}_{NaID}$ = 3$\times$EW$_{MgIb}$.   \cite{rup05} found that  most infrared luminous star forming galaxies below this line   have starburst driven  winds, while most galaxies above do not.

According to this basic analysis,  all  QSO2 in our sample but one (SDSS J1102+64) 
fall close to the stellar line,  indicating a small or null contribution of  interstellar absorption against the optical continuum in the central few kpc of the galaxy (but see \S5).  SDSS J1102+64, which shows evidence for strong interstellar absorption lies above the "lower line", that defines starburst driven winds.
The NaID doublet redshift is consistent within the large uncertainties with that expected from the narrow core of the [OIII] line. Thus, it does not show clear evidence for  a neutral outflow either. 

   \subsection{The ionization mechanism}

\begin{figure*}
\includegraphics{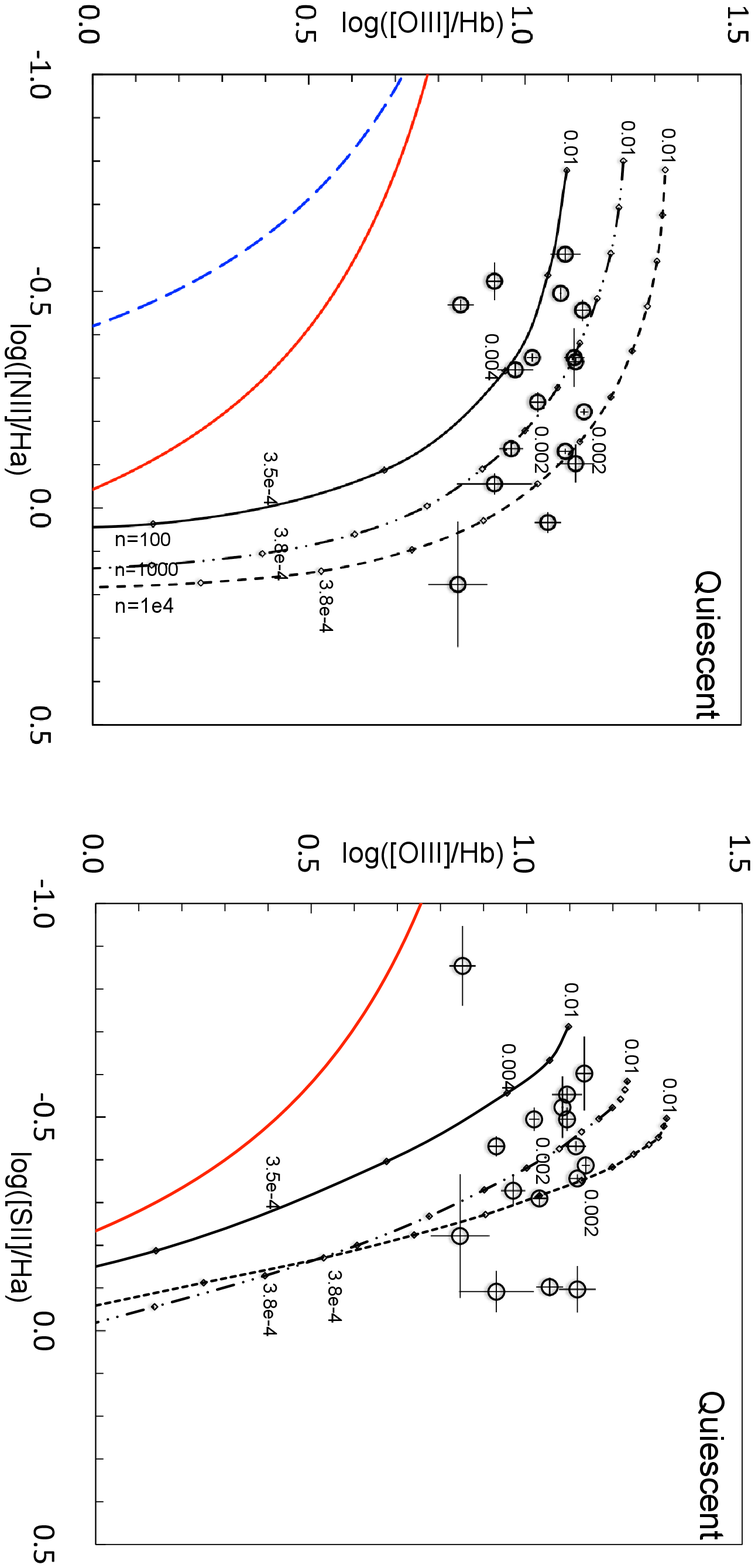}
\vspace{3.2in}
\includegraphics{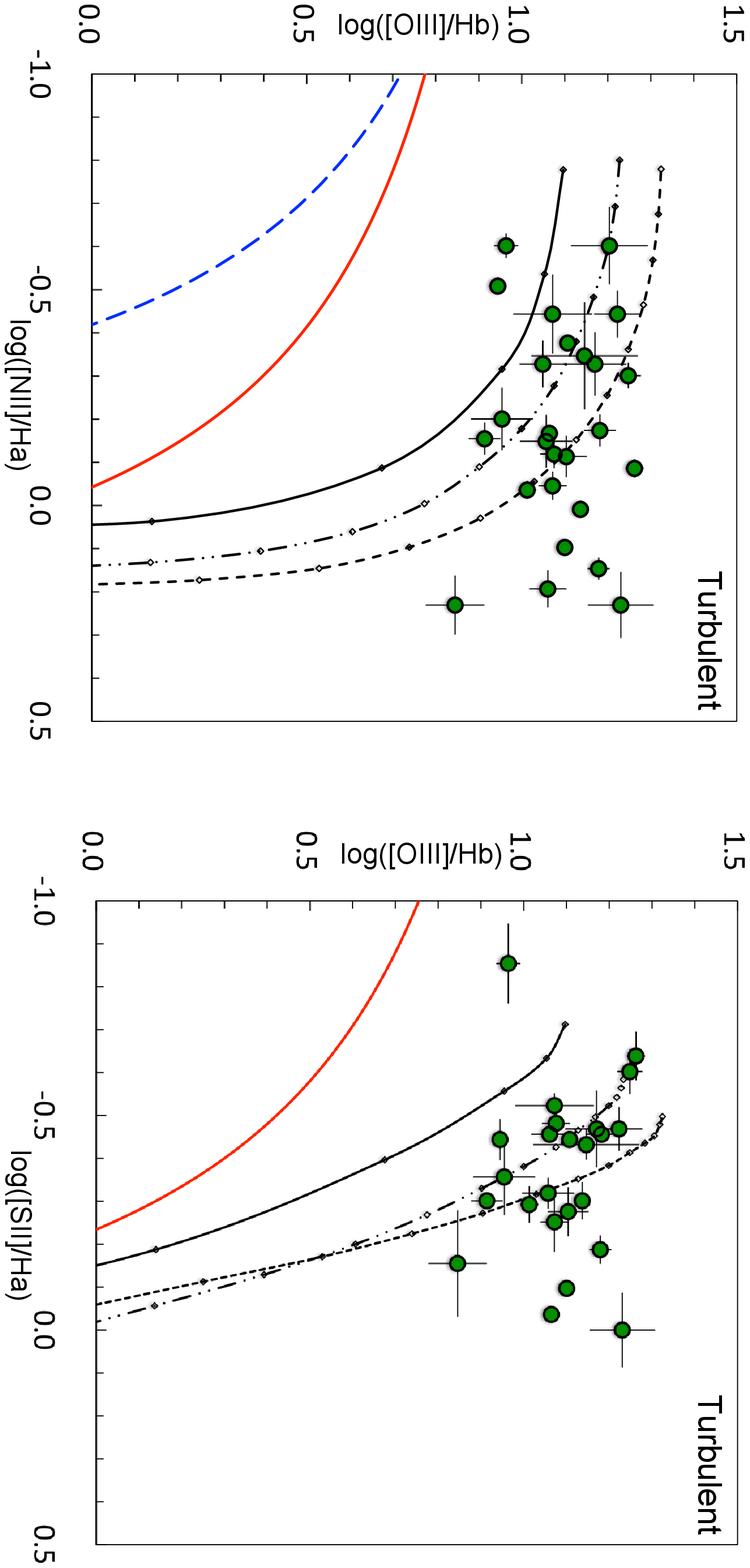}
\vspace{3.2in}
\caption{Diagnostic diagrams. The red solid and blue dashed lines are  the   AGN and starburst division lines by  
Kauffmann et al. (2003) and  Kewley et al. (2001) respectively. The black lines
are AGN photoionization model sequences (see text) for three gas density values  $n_H=$100 (solid), 10$^3$ (dot-dashed) and 10$^4$ cm$^{-3}$ (dashed).  The ionization parameter $U$  varies along each sequence (several $U$ values are
shown on the top diagrams).  All kinematic components (quiescent  and turbulent) occupy
the AGN area of the diagrams. This implies that both gaseous components are preferentially ionized by AGN related processes and not by stars. High densities in the range $\sim$100-10$^4$ cm$^{-3}$ are implied for most kinematic components by the models.}
\label{fig:Diag}
\end{figure*}

We show in Fig.\,\ref{fig:Diag} the diagnostic diagrams log(\oiiihb) vs. log(\niiha)  and log(\oiiihb) vs. log(\siiha) (Baldwin, Philips \& Terlevich \citeyear{bpt81}) with the location of
the quiescent (top) and the turbulent  (bottom) components.  Objects above the red solid lines  are AGN (Kewley et al. (\citeyear{kew01}).   Objects below  the dashed blue lines are starburst galaxies (Kauffmann et al. (\citeyear{kauf03}).
Objects in the intermediate area are composite in nature.

All turbulent and quiescent components  lie in 
the area of the diagrams occupied by AGN. Thus,  both the ambient and the outflowing gas are excited by AGN related
processes in all QSO2. 

The  black lines  represent three sequences of AGN photoionization models (Villar Mart\'\i n et al. \citeyear{vm08}) built with the multipurpose code MAPPINGS Ic (Ferruit et al. \citeyear{fer97})
The ionization parameter\footnote{$U=\frac{Q^{ion}}{4~\pi~r^2~n_H~c}$, where is $Q^{ion}$  the ionizing photon luminosity of the source, $r$ is the distance between the cloud and the ionizing source, $n_H$ is the hydrogen density at the illuminated face of the cloud and $c$ is the speed of light.} varies along each sequence
(some values are indicated on the top diagrams in Fig.\,\ref{fig:Diag}). The ionizing continuum
is a power law of index $\alpha$=1.5 ($F_{\nu} \propto \nu^{-\alpha}$), with a cut off energy  of 50 keV. The clouds  are considered to be  isobaric, plane-parallel, dust-free
ionization-bounded  slabs of density $n_H$ at the
illuminated face and characterized by solar abundances (Anders \& Grevesse \citeyear{anders89}).
Three different gas densities have been considered: $n_H$=100 (solid line), 10$^3$ 
(dot-dashed line) and 10$^4$ cm$^{-3}$ (dashed line). 

In general, the line ratios of both kinematic
components in most objects are consistent with  AGN photoionization models  (shock excitation related processes are not discarded, Allen et al. \citeyear{all08})\footnote{Some kinematic components show unusually high \nii/\ha~and \siiha~ratios, as can be appreciated in Fig.\,\ref{fig:Diag}. We notice that, in general, such extreme values affect both kinematic components, as well as the global line ratios (i.e. using the full line fluxes)  of a given object. Higher $n_H$ do not reproduce the large ratios, since collisional de-excitation of the \nii~and \sii~line becomes very efficient. These QSO2 might show some particular gas (e.g. high metallicity) and/or ionization continuum/mechanism properties. It is beyond the scope of this paper to explain the line ratios in these particular objects, and to  focus instead on the general trends.}.
Most data points span a range of model densities $n_H\sim$100-10$^4$ cm$^{-3}$.   A similar range of densities is known to exist in the narrow line region (NLR) of quasars and Seyferts (e.g. Bennert et al. \citeyear{ben06a}, \citeyear{ben06b}).  {\b It is probable that  higher density gas  also exists, since coronal lines are detected in  many of the QSO2 spectra. Lower densities are not discarded either.}

    \subsection{Electron densities and ionized gas masses}
  
  \subsubsection{Isolation of the ambient and outflowing components  in [SII]$\lambda\lambda$6716,6731} 
  
 We plan to constrain  the electron gas densities $n_e$\footnote{To a reasonable approximation: $n_e \sim n_H$ defined in \S4.4} of   the quiescent and turbulent components, by isolating them in the [SII]$\lambda\lambda$6716,6731 doublet (Osterbrock \citeyear{ost89}). 
 
\begin{figure*}
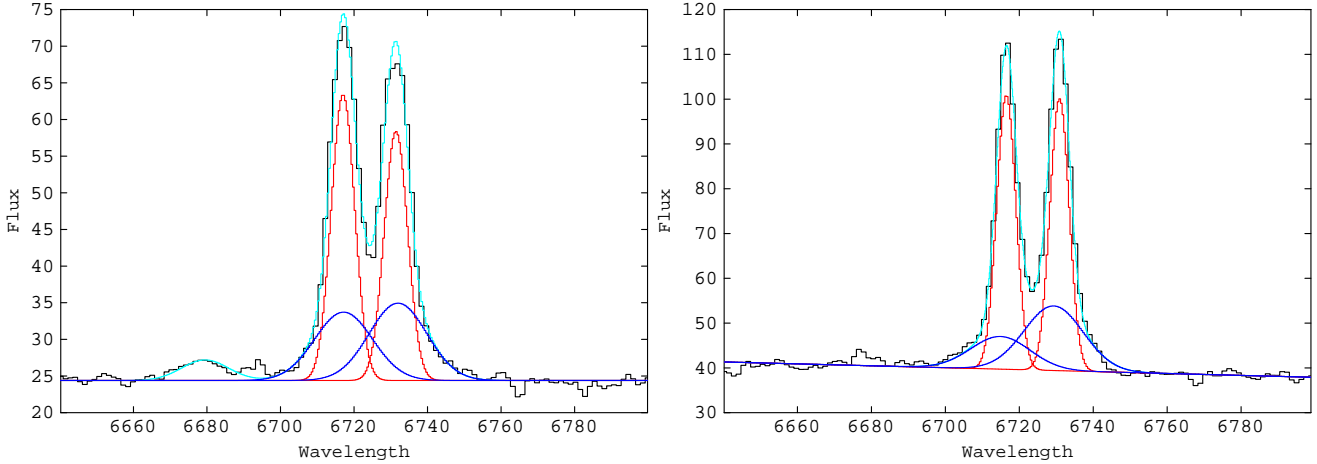

\includegraphics{fitsii0818+36.ps}
\includegraphics{fitsii1405+40.ps}
\vspace{2.5in}
\caption{Spectral fit of the [SII]$\lambda\lambda$6716,6731 doublet for SDSS J0818+36 (left) and SDSS J1405+34 (right). 
Color code as in Fig. 2. The fits suggest higher density in the outflowing gas. Flux is in units of 10$^{-17}$ erg s$^{-1}$ cm$^{-2}$.}
\label{fig:fitsii}
\end{figure*}

 Due to the complexity and broad FWHM of the spectral profiles, the doublet lines  are severely blended in most objects, and
 in general the fitting procedure does not give unambiguous results.   For this reason, we have focussed our analysis  on those QSO2 with the simplest spectral profiles (i.e. only 2 kinematic components as revealed by the fit
 of the [OIII] lines; see Table \,\ref{tab:tabfits}) and with no severe blending of the [SII] doublet. These are: SDSS J0818+36, SDSS J1300+54, SDSS J1405+34 and SDSS J2134+07 (e.g. Fig.\,\ref{fig:fitsii}).

  \begin{table*}
\centering
\begin{tabular}{llllllllllll} 
\hline
(1) & (2) & (3) & (4) & (5) & (6)  & (7) & (8) & (9) & (10) \\
Object &  $T_e$ & $\frac{F(\lambda6716)}{F(\lambda6731)}^{amb}$ &  $n_e^{amb}$ & $n_H^{[NII]/H\alpha}$  & $n_H^{[SII]/H\alpha}$  & $\frac{F(\lambda6716)}{F(\lambda6731)}^{outf}$ & $n_e^{outf}$ &  $n_H^{[NII]/H\alpha}$  & $n_H^{[SII]/H\alpha}$   \\
&	 K &  & cm$^{-3}$ & cm$^{-3}$ & cm$^{-3}$   &  & cm$^{-3}$   &cm$^{-3}$  &  cm$^{-3}$    \\ \hline
SDSS J0818+36 &   15700  & 1.12$^{+0.14}_{-0.18}$ & 410$^{+500}_{-240}$ & 10$^2$ & 10$^{2-3}$ & 0.81$^{+0.19}_{-0.24}$ & 1580$^{+4600}_{-870}$  & 10$^{3-4}$  & 10$^{3-4}$   \\
SDSS J1300+54 &  16970 &  1.15$^{+0.25}_{-0.20}$ & 360$^{+540}_{\rm low~n_e}$ & 10$^2$ & 10$^{2-3}$ & 0.98$^{+0.27}_{-0.31}$ & 800$^{+2400}_{-615}$ & 10$^{2-3}$ & 10$^{2-3}$   \\ 
SDSS J1405+34 &  16600 &  1.06$\pm$0.06 & 550$^{+165}_{-145}$  & 10$^2$ & 10$^{2-3}$ & 0.67$\pm$0.05 &   3200$^{1140}_{-750}$  & 10$^4$ & 10$^3$ \\
SDSS J2134+07 & 16740 & 0.94$\pm$0.04 & 940$^{+170}_{-150}$ & 10$^3$ & 10$^3$ &  0.86$^{+0.20}_{-0.26}$ &  1300$^{+3700}_{-740}$   & 10$^3$ & 10$^3$ \\ \hline
\end{tabular}
\caption{Comparison of $n_e$ for the ambient and outflowing components. (1) Object. (2) $T_e$ in K derived from the [OIII]$\lambda\lambda$4959,5007 and $\lambda$4363 integrated line fluxes. (3) [SII] doublet flux ratio for the ambient gas. (4)   $n_e$ implied by (3)  for the ambient gas. (5) and (6) show the range of $n_H$ (which is $\sim$$n_e$ to a reasonable approximation) suggested by the
location of the quiescent components in the \oiiihb~ vs. \niiha~  and \oiiihb~ vs. \siiha~diagrams. (7) to (10) are the same as (3) to (6) but for the outflowing gas.} 
\label{tab:dens1}
\end{table*}

To constrain the possible range of $\frac{F(\lambda 6716)}{F(\lambda 6731)}$  values we tried two methods:

1) free fits (i.e. no kinematic constraint from the [OIII] lines was applied).  For every individual kinematic component,  the   separation between
the doublet lines was fixed ($\Delta\lambda$=14.4 \AA, restframe) and they were forced to have the same FWHM.
In all 4 objects, it is found that the free fits require 2 kinematic components, as expected from the \oiii~lines. 

 2) fits with full kinematic constraints from the [OIII] lines. I.e. we assumed 2 kinematic components 
 with FWHM and velocity separation consistent with that inferred from   \oiii. Several 
 combinations of FWHM and velocity separation values were attempted for both kinematic components to account for the uncertainties of the \oiii~fits.

Different situations are found:  for SDSS J0818+36 the free fits produce unphysical results (e.g. $\frac{F(\lambda 6716)}{F(\lambda 6731)}>$2, Osterbrock \citeyear{ost89}) as well
as  very large errors.  Full kinematic constraints are required to
produce  successful fits. For SDSS J1300+54 and SDSS J2134-07, the free and fully constrained fits produce consistent results within the errors. 

For SDSS J1405+34, the free fit produces somewhat different results than expected from the [OIII] kinematic substructure. Two kinematic components 
are isolated with FWHM=200$\pm$10 \kms~ and 630$\pm$40 \kms~respectively, and a velocity separation $\Delta v=$-10$\pm$10 \kms~ of the broad relative to the narrow component (vs. FWHM=270$\pm$11 \kms~ and 830$\pm$19 \kms~  and $\Delta v$=-76$\pm$10 \kms measured in the [OIII] lines). 
 We do not discard a physical origin for these differences, since the [OIII]  and the [SII] lines  might map somewhat different gas regions due  to reddening, for instance.
 
While  $\frac{F(\lambda 6716)}{F(\lambda 6731)}$  is in general well constrained for
the quiescent (i.e. ambient) component, it is much more uncertain for the turbulent (i.e. outflowing) component. The main sources of uncertainty are the  level and slope  of the continuum.  All uncertainties considered, the  $\frac{F(\lambda 6716)}{F(\lambda 6731)}$ values and errors quoted 
  in Table \,\ref{tab:dens1} account for the full range of successful (and physically meaningful)   fits resulting from applying the diversity of methods described above.

  \subsubsection{Electron densities} 

 The $n_e$ inferred from the [SII] line ratios depends on the electron temperature $T_e$ (Osterbrock \citeyear{ost89}), with $n_e(T) = n_e([SII]) \times \sqrt{T_e/10000}$. This has been estimated for each object using the [OIII] $\lambda\lambda$5007,4959 and $\lambda$4363 integrated line fluxes (Table \,\ref{tab:dens1}). Although it is not possible to measure $T_e$ for both kinematic components,  the dependence of $n_e$ would  be only weak within the $T_e$ range they are expected to span. 
  
  We show   $n_e$  in Table \,\ref{tab:dens1} for the ambient (column 4) and outflowing (column 8) components in the 4 QSO2 where [SII] could be fitted.

  We also show the range of densities  of the  AGN photoionization models  closest to the location of the data in the diagnostic diagrams in Fig.\,\ref{fig:Diag}.  
    In spite of the large uncertainties on $n_e$, we find in the 4 QSO2 that the outflowing gas has higher $n_e$ (often $>$10$^3$ cm$^{-3}$) 
than the ambient gas (in general, $n_e\sim$few$\times$hundred -10$^3$ cm$^{-3}$). 
This is implied both by the $\frac{F(\lambda 6716)}{F(\lambda 6731)}$ ratios and   the location of the data in the
  diagnostic diagrams relative to the AGN photoionization models.  The higher \oiiihb~ shown by the turbulent components compared to the quiescent gas (see \S4.2)
  can now be explained as a consequence of the difference in $n_e$.

The inferred $n_e$  are not expected to be representative of the whole gaseous regions, where a density gradient is expected to occur.
The results above prove the existence of a high density ($\ga$1000 cm$^{-3}$)  in the outflowing gas, higher than that of the ambient gas.
      
 \section{Discussion}

\subsection {\it Ubiquity of ionized outflows in optically selected QSO2}

 We have investigated the existence and properties of nuclear ionized outflows in the 21 most nearby QSO2 at $z\la$0.1,  by means of the
spectral decomposition of the main optical emission lines, using their SDSS optical spectra.
The turbulence parameter $R=\frac{\rm FWHM_{[OIII]}}{\rm FWHM_{\star}}$  provides the basis for an efficient discrimination method between the turbulent ($R>$1.4)
and the quiescent ($R\le$1.4) gaseous components.  Based on their kinematic properties we identify the turbulent components  with emission from outflowing gas and  the quiescent components with ambient gas  which has not been reached by the outflows. 

We have found ionized outflows in  all but one object (20/21). This reinforces the
 result  by \cite{vm11b} for QSO2 at $z\sim$0.3-0.6, that ionized outflows are ubiquitous in the NLR of optically selected QSO2 (see also Liu et al. \citeyear{liu13b}).

\subsection  {\it The outflowing gas is patchy and/or asymmetric}

The outflowing gas shows a preference for blueshifts relative to $z_{sys}$, indicating differential reddening by dust. Based on the Balmer decrement we find that the outflowing gas is more heavily reddened than the ambient gas (see also Holt et al. \citeyear{hol11}). This supports that dust extinction severely affects  its emission. 

In several QSO2 both the receding and the approaching parts of the outflow have been detected. The fact that the redshifted outflowing gas can be brighter than the blueshifted  one (e.g. SDSS J0802+25, SDSS J1455+32, see Table \,\ref{tab:tabfits}) suggests that the reddening is patchy  and/or the outflow gas distribution is asymmetric.  It is not possible to disentangle whether the dust is mixed with the outflowing gas or  it lies in an intervening structure that obscures the outflow more than the ambient, more extended gas.

\subsection {\it Spatial location and gas density of the outflows}

The outflowing gas shows optical line ratios consistent with AGN  (rather than stellar) ionization related mechanisms (\S4.4), implying that it is located
  inside the ionization cones of the quasar (see also \cite{vm11b}). This is further supported by the smaller $|V_S|$ values for the outflowing gas for QSO2 compared with QSO1. As an example, about half of the objects in the heterogenous sample
 of QSO1 at $z\la$0.2 studied by \cite{rose13}  show $|V_S|\ga$200 \kms, while only 5 QSO2 in our sample exceed this value. This is consistent with the orientation based unification scenario: since the cones axis is close to the line of sight in QSO1, larger $|V_S|$ values are expected provided the outflows expand within the ionization cones.

  The outflows we see are  constrained within a radial distance from the AGN of $r<$3 kpc, given the 3" SDSS fiber aperture diameter. They  are therefore located in the NLR.

We have found that the
 outflowing gas has a higher density than the ambient gas with  
$n_e\ga$1000 cm$^{-3}$ and  $n_e\sim$few$\times$hundred -10$^3$ cm$^{-3}$ respectively (Table \,\ref{tab:dens1}). 
  $n_e >$ several$\times$10$^3$ cm$^{-3}$ have been measured in prominent ionized quasar outflows detected in absorption (e.g. Arav et al. \citeyear{arav13}).
  Density enhancement of the outflowing gas  has been found in some radio galaxies (e.g.  Holt et al.  \citeyear{hol11},  Villar Mart\'\i n et al. \citeyear{vm99}) and local Ultraluminous Infrared Galaxies (ULIRG, Arribas et al. \citeyear{arr14}).    In particular, \cite{hol11} infer   $n_e\sim$10$^4$-several$\times$10$^5$ cm$^{-3}$ for the outflowing gas in the compact radio source  PKS 1345+12 at $z=$0.12.

 Based on studies of the radial dependence of  the NLR density in Seyfert 1 and Seyfert 2 galaxies ($n_e \propto r^{-1.14 \pm 0.1}$,  Bennert et al. \citeyear{ben06a}, \citeyear{ben06b}), it is reasonable to expect the density to decrease at increasing distance $r$ from the central AGN also in quasars. If such is the case, the enhanced $n_e$ of the outflowing gas  can be explained if  it  
is concentrated in a smaller region closer to the AGN, while the ambient gas extends further.    Another possibility is  that the shocks induced during the passage of the outflow front have produced the compression of the gas.  

We propose the first scenario to be more plausible. We warn the reader that the argumentation that follows is valid provided photoionization by the quasar is the dominant ionization mechanism of  the outflow compared to shock excitation.
Let us focus on the 4 QSO2 where  $n_e^{outf}$ and $n^{amb}$ could be
determined using [SII] (Table \,\ref{tab:dens1}).

 Using the definition of the ionization parameter $U=\frac{Q^{ion}}{4~\pi~r^2~n_e~c}$ (\S4.4) and since $Q^{ion}$   is the same for both gaseous components, then 
$\frac{r^{amb}}{r^{outf}}=\sqrt{\frac{U^{outf} ~\times~ n_e^{outf}}{U^{amb} ~\times~ n_e^{amb}}}$.
$U^{outf}$ and $U^{amb}$ are inferred for each object in Table 5
from the photoionization models closest to the location of the data in the \niiha~vs. \oiiihb~diagram (Fig.\,\ref{fig:Diag}). We find that in all cases  
$U^{outf} \sim U^{amb}$ and $\frac{n_e^{outf}}{n_e^{amb}}\sim$1.4-5.8 (Table 6) depending on the objects. Thus, the outflowing gas is located  $\sim$1.2-2.4 times closer to the central engine than the ambient gas.
Since this  is expected to extend at least across the fiber diameter  (Liu et al. \citeyear{liu13a}), then $r^{outf}\sim$1-2 kpc depending on the object\footnote{The ambient gas possibly extends  $\ga$15 kpc from the AGN, but notice that the $U$ and $n$ ratios refer only to the gas within the fiber aperture.}.
In spite of the large uncertainties involved in our calculations, these values are in  good agreement with the  wind radii $r\sim$2-5 kpc  measured by \cite{rup13}  for the three  QSO in their sample (see also Arav et al. \citeyear{arav13}).  

We can further test whether the $n_e^{outf}$ and $r^{outf}$ ranges are reasonable, by estimating the implied  $Q^{ion}$ using again the definition of $U$. Assuming $r^{outf}\sim$2 kpc, $Q^{ion}\sim$(0.7-2.0)$\times$10$^{56}$ s$^{-1}$ in all cases, which is consistent with  quasar photon ionizing luminosities (e.g. Yee \citeyear{yee80}).

We thus propose that  the ionized outflows are  concentrated  in a smaller region ($r^{outf}\sim$1-2 kpc) than the ambient gas and closer to the central engine.  The stratification of gas density of the NLR would explain the higher $n_e$ of the outflowing gas. It 
 may also result in the more central regions being  more heavily obscured, in coherence with our finding that  the outflowing gas is more  reddened (\S4.2; see also Holt et al. \citeyear{hol11}).

\subsection {\it Outflow  masses and mass flow rates}

We have calculated the mass of the ambient and outflowing gas for the 4 QSO2  in Table \,\ref{tab:dens1} as  $M = \frac{m_p}{\alpha^{eff}_{H\beta}~h\nu_{H\beta}}\frac{L(H\beta)}{n_e}$ where $m_p$ is the proton mass, $\alpha^{eff}_{H\beta}$ is the H$\beta$ effective recombination coefficient (case B) and $h\nu_{H\beta}$ is the energy of an H$\beta$ photon (Osterbrock \citeyear{ost89}).
For each kinematic component, $L(H\beta)$ has been  corrected for reddening using the corresponding Balmer decrement $\frac{H\alpha}{H\beta}$ (Table \,\ref{tab:tabfits}). The results are shown in Table \,\ref{tab:dens2}.

In these 4 objects, $\sim$7-26\% of the ionized gas mass in the inner $r<$3  kpc is involved in the ionized outflow. Outflow
masses in the range $M_{outf}\sim$(0.3-2.4)$\times$10$^6$ \msun~are inferred for the density values  $n_e^{outf}$ in Table \,\ref{tab:dens2}. Using the minimum possible $n_e^{outf}$ allowed by the uncertainties (Table \,\ref{tab:dens1}) would 
result in $M_{outf}\sim$10$^6$-10$^7$ \msun.  For comparison, molecular outflows in quasars are capable of dragging masses $>$10$^8$ \msun~ at rates ${\dot M}\ga$several$\times$100 \msun~ yr$^{-1}$
  (e.g. Cicone et al. \citeyear{cic13}). 

The mass outflow rate  can be calculated approximately as ${\dot M} \sim \frac{v_0~M^{outf}}{r^{outf}}$.
 Assuming a spherical geometry,  $v_0$ is the expanding velocity of the outflowing bubble and $r$ its radius. 	 $v_0$  is likely to be $>$$V_s$ (Table 2), due to projection effects.  \cite{liu13b} estimate a median  velocity $\sim$760 \kms~ for the ionized outflows in their sample of QSO2. Ionized outflow velocities $\la$1000 \kms~ are measured by \cite{rose13} in their sample
 of QSO1 at $z\la$0.2. Values as high as $\sim$3000 \kms~ have been proposed in  some extreme cases (Rupke \& Veilleux \citeyear{rup13}). Thus, by considering $v_0=$3000 \kms~we will obtain an upper limit on ${\dot M}$. For $r^{outf}\sim$2 kpc (see \S5.3),  
  ${\dot M}<$4 M$_{\odot}$ yr$^{-1}$ in the 4 QSO in Table 5. Although our calculations are  rather uncertain, the upper limits are consistent  with the  mass outflow rates of the ionized outflows derived for the quasars in  \cite{rup13}  sample.   Our ${\dot M}$ are very much below   ${\dot M}\ge$ 2000 M$_{\odot}$ yr$^{-1}$ proposed by \cite{liu13b}. The main reason for the discrepancy is the much lower density adopted by these authors.

 The calculations above are simplistic and illustrative, since the outflow gas is expected to have a gradient in density and velocity (a valuable step forward on this regard has been performed by Liu et al. \citeyear{liu13b}).
Nevertheless, a useful conclusion can be extracted from our analysis:   the ionized outflows in the NLR of QSO2 contain very high density gas ($\ga$1000 cm$^{-3}$), which  contributes the bulk of the NLR
line luminosities.  Similar calculations as presented above are often found in the literature
assuming gas densities $n<$10 cm$^{-3}$. This can lead to overestimations of gas masses and mass outflow rates of a factor $>$100. 
Outflow masses and rates  $\sim$100 higher would require the existence of a
large reservoir of low density outflowing gas. This is not discarded but it would be undetectable in the spatially integrated spectrum,  due to the
relatively faint line luminosities.

 \begin{table}
\centering
\tiny
\begin{tabular}{lllllllll} 
\hline
(1) & (2) & (3) &  (4)   & (5) & (6)  \\ 
Object &  $n_e^{amb}$ &   $M_{amb}$ &  $n_e^{outf}$  & $M_{outf}$ & $M(\%)_{outf}$  \\ 
		& cm$^{-3}$  & $\times$10$^6$ \msun & cm$^{-3}$	&    $\times$10$^6$  \msun  \\ \hline
SDSS J0818+36 &  410 &   6.9  & 1580 & 0.5  & 7\% \\
SDSS J1300+54 & 360 & 7.3 & 800 & 2.4   & 25\%  \\
SDSS J1405+40 &  550 & 2.3  & 3200 & 0.8 & 26\%   \\
SDSS J2134+07 & 940 &  0.9 & 1300 & 0.3  & 25\%  \\
\hline
\end{tabular}
\caption{Mass  of the ionized ambient (3) and outflowing gas (5) using the $n_e$ (columns 2 and 4) constrained  
with the [SII] lines (Table \,\ref{tab:dens1}). Column (6) quotes the percentage of the total ionized gas mass involved in the outflow.} 
\label{tab:dens2}
\end{table}

 \subsection  {\it  The triggering mechanism}

Based on the $q$ parameter (Fig.\,\ref{fig:FIRradio}) and its relation with $R_{max}$, we proposed in \S4.3 that the most extreme outflows
($R_{max}\ge$4.7, Group 1) are generated by radio structures. These must be constrained within  spatial scales $\la$5 kpc, given the size of the 3" fiber.
 In addition to the different behavior with $q$,  objects in Groups 1 and 2 also differ clearly in the $R2$ vs. $R_{max}$ (Fig.\,\ref{fig:R_radio}, left) diagram. They define 2 separate approximately parallel anticorrelations. At $R_{max}=4.7$ a sudden change is found such that $R2$ becomes much higher than expected from the general trend defined by objects with
lower turbulence. This suggests that a different mechanism is working in  Groups 1 and 2.

 The less extreme outflows ($R_{max}\sim$2-3, Group 2) are likely to have a different origin.
  This is likely to be related to the quasar activity as well (e.g. an accretion disk wind, see \S1),  since  the outflowing gas is  located inside the ionization cones in the NLR and  excited by AGN related  processes.  The high density of the outflowing gas ($n_e\ga$1000 cm$^{-3}$) is also more typical of the dense AGN NLR environments, rather than starburst driven winds ($\sim$several$\times$100 cm$^{-3}$; e.g. Arribas et al. \citeyear{arr14}).
  The larger $R2$ for a given $R$ in Group 1 compared to Group 2 are naturally explained in our scenario. Since the radio structures are much more efficient at inducing extreme outflows,   they are possibly also capable of dragging and accelerating  a given amount of mass to more extreme velocities.

In both groups, $R2$ decreases with $R$.   Given the density enhacement of the outflowing gas, this suggests that smaller gas mass fractions 
 are involved in the outflow as the turbulence increases (see also \cite{vm11b}).  This is independent of the outflow triggering mechanism.

The presence of stellar winds cannot be  discarded, but  they are not expected to play a significant role in driving the ionized outflows in the QSO2 sample. Different works suggest that AGN are capable of triggering faster and more energetic outflows than pure starbursts. Even in ULIRG-AGN systems, where star formation is extremely active,  the AGN seems to be the main driver of the ionized winds (Rodr\'\i guez Zaur\'\i n et al. \citeyear{rod13},   Rupke \& Veilleux \citeyear{rup13}).

Thus, we propose that there are two dominant outflow mechanisms in this QSO2 sample: radio jet-induced (Group 1) and another AGN related mechanism (e.g. accretion disk wind) (Group 2). Both might
be present in all QSO2 at some level, but the effect of the radio jets becoming dominant is obvious both in a more extreme gas
turbulence and the larger fraction of gas mass involved in the outflow.  The different directionality of
 the radio jet and the disc wind modes can have important implications on  the impact these feedback mechanisms have
 on the surrounding medium at different spatial scales.

 To test our interpretation we predict  that high resolution radio maps will reveal the presence of jets in those objects with the most extreme outflows (Group 1)\footnote{This is certainly the case for SDSS J1440+10 (Heckman et al. \citeyear{hec97}). The 8.4GHz VLA radio continuum map (0.26" resolution) shows a triple radio source. 
The correlation between the radio and [OIII] morphologies  proves the impact  of the radio source on the NLR.}.
 Because the radio-jet induced outflows are expected to be much more collimated than the wide angle disk winds,   the ionized nebulae are likely to be  more highly collimated and show more turbulent kinematics in  Group 1 compared to Group 2. If the outflows expand to large distances  $>$several kpc from the AGN (e.g. Liu et al. \citeyear{liu13b}, Humphrey et al. \citeyear{hum10}), the nebular morphology 
could be characterized across scales of several arcsec and the differences should be obvious. 
Indeed, such a difference between radio quiet and radio intermediate/loud QSO2 has already been hinted by \cite{liu13a}.

 \vspace{0.2cm}

 \subsection  {\it  Neutral winds}
\vspace{0.2cm}

Our basic analysis  reveals no clear evidence for a significant component of neutral ISM  absorption and consequently, 
neither for neutral outflows in QSO2. The only exception is SDSS J1102+64, where ISM absorption is  strong. However, neither in this case  it is clearly associated with a neutral wind.

It has to be noted that NaID absorption must be traced against regions with a bright stellar continuum. In cases where the AGN is responsible for driving turbulent outflows, the region under study may be much more localized and/or shifted spatially from stellar regions. Other methods are required to study the neutral gas component involved in such systems \citep[e.g.][]{mor05,mor13,mah13}. It is also possible that the ISM has been so highly ionized by the quasar
that the neutral component is relatively much less significant (even undetectable) than in starburst galaxies (e.g. Rupke, Veilleux \& Sanders \citeyear{rup05}). This is actually suggested by
the null or weak  ISM contribution to the NaID feature.

 \section{Summary and conclusions}
 
 We have investigated the existence and properties of nuclear ionized outflows in the 21 most nearby QSO2 at $z\la$0.1,  by means of the
spectral decomposition of the main optical emission lines in their SDSS spectra. Most objects are radio quiet according to the 
radio to [OIII] luminosity ratio. A few are radio-intermediate, but in all cases they have radio luminosities well below typical radio loud AGN.

The turbulence parameter $R=\frac{\rm FWHM_{[OIII]}}{\rm FWHM_{\star}}$  provides the basis for an efficient discrimination method between the turbulent ($R>$1.4) 
and the quiescent ($R\le$1.4) gaseous components.  Based on their kinematic properties we identify the turbulent components  with emission from outflowing gas and  the quiescent components with ambient gas  which has not been reached by the outflows. 
We have detected ionized outflows in all  but one object. Thus, the detection
 rate is 95\%. This result extends to the lowest $z$ the
 conclusion by \cite{vm11b} for QSO2 at $z\sim$0.3-0.6, that ionized outflows are ubiquitous in optically selected obscured quasars.

 The ionized outflows are located in the narrow line region. Both the ambient and the outflowing gas are ionized by AGN related processes. The outflowing gas is more highly reddened than the ambient gas   (\ha/\hb= 4.91$\pm$0.03 and  3.68$\pm$0.02 respectively). It is also  denser  $n_e\sim$few$\times$hundred -10$^3$ cm$^{-3}$ for the ambient gas, while $n_e\ga$1000 cm$^{-3}$ is frequent in the outflowing gas. To explain these results, we propose that the bulk of the outflow line emission
 is originated in a  more compact region ($r\sim$1-2 kpc) and closer to the AGN than the ambient gas. 

 Such high densities imply low $M_{outf}\sim$(0.3-2.4)$\times$10$^6$ \msun~ and mass outflow rates $\dot M<$few \msun~ yr$^{-1}$. Larger values could exist if most of the outflow mass is contained in a reservoir of
  low density gas ($n_e<$few  cm$^{-3}$). This  might dominate 
 the ionized outflow mass, but  not the line luminosity which would be too faint to detect in spatially integrated spectra.

The triggering mechanism of the outflows is related to the nuclear activity, rather than  star formation.
The QSO2 can be classified in two clearly differentiated groups according to the  behavior and properties of the outflowing gas.
QSO2 in Group 1 (5 objects) show the most extreme turbulence ($R_{max}\ge$4.7), they have on average higher radio luminosities and lower  $q$ parameter, which parametrizes the excess of radio emission or radio-loudness.  In this group, it is found that the lowest $q$ values (i.e. larger radio excess) are associated with the  most extreme gas turbulence (larger $R_{max}$).   QSO2 in Group 2 (15 objects) show less extreme turbulence  ($R_{max}\sim$2-3), they have lower radio luminosities and, on average, higher $q$ values.

Based on these results, we propose that two competing 
outflow mechanisms are at work:  radio jets and another AGN related process  (possibly accretion disk winds).  
Although both mechanisms might be present in at least some objects,
the radio jet induced outflows are dominant in Group 1 (5/20 QSO2), while the disk wind dominates in Group 2 (15/20 QSO2).  In this scenario, we  find that the radio jet
mode is  capable of producing  more extreme (turbulent) outflows. 
Independently of the outflow origin, our results suggest that the larger the turbulence, the smaller the mass fraction involved in the outflow, in coherence with \cite{vm11b}. 

To test this interpretation we predict  that high resolution VLBA imaging will reveal the presence of jets in Group 1 QSO2. If the outflows expand to  distances  $>$several kpc from the AGN, the
 morphologies and kinematic properties could be characterized across scales of $\ga$several arcsec at $z\sim$0.1. 
We predict that the morphology of the extended ionized nebulae is more highly collimated and kinematically perturbed
 for objects in Group 1.  

 Our basic analysis  reveals no  evidence for  neutral outflows in QSO2.  The ISM might be too highly ionized across large spatial scales
 (indeed  no significant component of neutral ISM  absorption   is detected).  Alternatively, the detection method,  consistent of tracing NaID absorption   against regions with a bright stellar continuum, may be
 inadequate for tracing AGN induced outflows.

\section*{Acknowledgments}

Special thanks to Santiago Arribas for useful discussions and comments on the manuscript.

This work has been funded with support  from the Spanish Ministerio de Econom\'\i a y Competitividad through  the
grant AYA2010-15081, AYA2012-32295 and AYA2010-21161-C02-01.

The work is based on data from Sloan Digital Sky Survey. Funding for the SDSS and SDSS-II has been provided by the Alfred P. Sloan Foundation, the Participating Institutions, the National Science Foundation, the U.S. Department of Energy, the National Aeronautics and Space Administration, the Japanese Monbukagakusho, the Max Planck Society, and the Higher Education Funding Council for England. The SDSS Web Site is http://www.sdss.org/.

The SDSS is managed by the Astrophysical Research Consortium for the Participating Institutions. The Participating Institutions are the American Museum of Natural History, Astrophysical Institute Potsdam, University of Basel, University of Cambridge, Case Western Reserve University, University of Chicago, Drexel University, Fermilab, the Institute for Advanced Study, the Japan Participation Group, Johns Hopkins University, the Joint Institute for Nuclear Astrophysics, the Kavli Institute for Particle Astrophysics and Cosmology, the Korean Scientist Group, the Chinese Academy of Sciences (LAMOST), Los Alamos National Laboratory, the Max-Planck-Institute for Astronomy (MPIA), the Max-Planck-Institute for Astrophysics (MPA), New Mexico State University, Ohio State University, University of Pittsburgh, University of Portsmouth, Princeton University, the United States Naval Observatory, and the University of Washington.

\end{document}